\documentclass[journal]{IEEEtran}
\IEEEoverridecommandlockouts
\usepackage{cite}
\usepackage{amsmath,amssymb,amsfonts}
\usepackage{hyperref}
\usepackage{algorithmic}
\usepackage{graphicx}
\usepackage{textcomp}
\usepackage{xcolor}
\usepackage{adjustbox}
\usepackage{verbatim}
\usepackage{multirow}
\usepackage{booktabs}
\usepackage{algorithm2e}
\usepackage{orcidlink}
\usepackage{threeparttable}

\usepackage{xcolor}
\hypersetup{
    colorlinks,
    linkcolor={black},
    citecolor={black},
    urlcolor={blue!80!black}
}

\def\BibTeX{{\rm B\kern-.05em{\sc i\kern-.025em b}\kern-.08em
    T\kern-.1667em\lower.7ex\hbox{E}\kern-.125emX}}
    
\begin{document}
\bstctlcite{IEEEexample:BSTcontrol}

\title{A Flexible Template for Edge Generative AI with High-Accuracy Accelerated Softmax \& GELU \\
%
}

\author{Andrea~Belano\orcidlink{0009-0005-8576-3502},
Yvan~Tortorella\orcidlink{0000-0001-8248-5731},
Angelo~Garofalo\orcidlink{0000-0002-7495-6895},~\IEEEmembership{Member,~IEEE,}
Luca~Benini\orcidlink{0000-0001-8068-3806},~\IEEEmembership{Fellow,~IEEE,}
Davide~Rossi\orcidlink{0000-0002-0651-5393},~\IEEEmembership{Senior~Member,~IEEE,}
Francesco~Conti\orcidlink{0000-0002-7924-933X},~\IEEEmembership{Senior~Member,~IEEE}%
\thanks{This work was financially supported in part by the Chips-IT Foundation, in part by the ChipsJU project ISOLDE (g.a. 101112274), and in part by the Spoke 1 on Future HPC of the Italian Research Center on High-Performance Computing, Big Data and Quantum Computing (ICSC) funded by MUR Mission 4 - Next Generation EU. \textit{(Corresponding author: Andrea Belano.)}}%
\thanks{A. Belano, A. Garofalo, L. Benini, D. Rossi, and F. Conti are with the Department of Electrical, Electronic, and Information Engineering (DEI), University of Bologna, 40126 Bologna, Italy; e-mail: andrea.belano2@unibo.it.}%
\thanks{A. Belano is also with the University of Pavia, 20100 Pavia, Italy.}%
\thanks{Y. Tortorella performed this work while at DEI, University of Bologna, Italy; he is currently with Eggtronic, 41126 Modena, Italy.}%
\thanks{A. Garofalo and L. Benini are also with ETH Z\"urich, 8092 Z\"urich, Switzerland.}
}

\markboth{Pre-Print submitted to IEEE Journal of Emerging and Selected Topics in Circuits and Systems}%
{Belano \MakeLowercase{\textit{et al.}}}

\maketitle

\begin{abstract}
Transformer-based generative Artificial Intelligence (GenAI) models achieve remarkable results in a wide range of fields, including natural language processing, computer vision, and audio processing. However, this comes at the cost of increased complexity and the need of sophisticated non-linearities such as softmax and GELU. Even if Transformers are computationally dominated by matrix multiplications (MatMul), these non-linearities can become a performance bottleneck, especially if dedicated hardware is used to accelerate MatMul operators.
In this work, we introduce a GenAI BFloat16 Transformer acceleration template based on a heterogeneous tightly-coupled cluster containing 256KiB of shared SRAM, 8 general-purpose RISC-V cores, a 24$\times$8 systolic array MatMul accelerator, and a novel accelerator for Transformer softmax and GELU non-linearities: SoftEx.
SoftEx introduces an approximate exponentiation algorithm balancing efficiency (121$\times$ speedup over glibc's implementation) with accuracy (mean relative error of 0.14\%).
In 12nm technology, SoftEx occupies 0.039 mm$^2$, only 3.22\% of the cluster, which achieves an operating frequency of 1.12 GHz. 
Compared to optimized software running on the RISC-V cores, SoftEx achieves significant improvements, accelerating softmax and GELU computations by up to 10.8$\times$ and 5.11$\times$, respectively, while reducing their energy consumption by up to 10.8$\times$ and 5.29$\times$. These enhancements translate into a 1.58$\times$ increase in throughput (310 GOPS at 0.8V) and a 1.42$\times$ improvement in energy efficiency (1.34 TOPS/W at 0.55V) on end-to-end ViT inference workloads.


\end{abstract}

\section{Introduction}
Transformers~\cite{vaswaniAttentionAllYou2017} are the main deep neural network (DNN) models driving the evolution of modern Artificial Intelligence (AI), due to their groundbreaking performance in a wide range of fields, including both \textit{perceptive} tasks such as computer vision~\cite{liuSwinTransformerHierarchical2021,jiCompressedDomainVisionTransformer2024} and audio processing~\cite{gongASTAudioSpectrogram2021} as well as \textit{generative} AI (GenAI) applications, such as natural language processing~\cite{touvronLLaMAOpenEfficient2023,radfordLanguageModelsAre2019} and video processing~\cite{duCGVCTContextualGenerative2024,yanFVIFormerFlowGuidedGlobalLocal2024,zhuTMGANTransformerBasedMultiModal2024}.
While the State-of-the-Art for Transformer-based GenAI models is based on cloud inference of models in the order of $\sim10^{11}$ parameters in data centers, there is a significant interest in enabling the execution of smaller models ($\sim10^8$--$10^9$ parameters) directly at the edge, to enable low-latency applications (e.g., robotics~\cite{brohanRT1RoboticsTransformer2023}), reduce wireless traffic congestion, and improve security and privacy in edge GenAI applications~\cite{aleEmpoweringGenerativeAI2024,wangOverviewGenerativeAI2023}.

Compared to previous-generation AI algorithms, such as Convolutional Neural Networks (CNNs), most Transformers (both for perceptive and generative tasks) typically feature a larger size and a more complex architecture.
For example, for similar tasks ResNet employed $\sim$25 million parameters~\cite{heDeepResidualLearning2016} (and even less for mobile-oriented CNNs such as MobileNets~\cite{howardMobileNetsEfficientConvolutional2017}), versus up to 632 million of the original Vision Transformer (ViT)~\cite{dosovitskiyImageWorth16x162020}; moreover, each layer of ViT is more complex than a ResNet convolutional layer, comprising multi-head self-attention (MHSA) and additional projections.
Despite several works in this direction~\cite{dettmersQLoRAEfficientFinetuning2023,frantarGPTQAccuratePostTraining2023,linAWQActivationawareWeight2024,xiaoSmoothQuantAccurateEfficient2023}, the full-integer execution of Transformers has proven more elusive than that of CNNs, mostly because many Transformer-based models are very expensive to train and/or trained on huge non-public datasets (or with human feedback) and, therefore, are not well-suited to quantization-aware fine-tuning~\cite{frantarGPTQAccuratePostTraining2023,xiaoSmoothQuantAccurateEfficient2023}.
Their large scale implies that the deployment of Transformer-based GenAI on edge devices is essentially impossible without the aid of hardware acceleration.

The vast majority of operations in a Transformer is linked to matrix-multiplications (MatMuls) -- either between dynamic inputs and static weights, such as in the case of Projections and Feed-Forward Network (FFN) layers, or between dynamic heads within the Attention layer in MHSA.
Therefore, Transformer accelerators dedicate most of their silicon budget to \textit{tensor processing units} to accelerate linear layers~\cite{wieseAttentionbasedTinyMLHeterogeneous2024,keller17956TOPSDeep2022}.
Multi-core clusters with tensor processing units have recently been proposed as an architectural template that can be used to accelerate Transformers in a low-power edge setting~ \cite{schererDeeployEnablingEnergyEfficient2024,wieseAttentionbasedTinyMLHeterogeneous2024,tortorellaRedMuleMixedprecisionMatrix2023} (with quantized networks), but potentially scalable to larger devices capable of targeting complete edge GenAI workloads~\cite{paulinOccamy432Core2812024}.
In this work, our overall goal is to design a multi-core heterogeneous cluster template tuned for the execution of GenAI without implying aggressive quantization (i.e., running at the native BFloat16 precision of Transformers), but still achieving throughput and efficiency compatible with edge-based GenAI.
As a baseline for our work, we employ an 8-core RISC-V cluster, based on the open-source PULP\footnote{\url{https://pulp-platform.org}} template, integrating a tensor processing unit based on the RedMulE architecture~\cite{tortorellaRedMuleMixedprecisionMatrix2023}.

Heterogeneous software/hardware computing provides the necessary flexibility to run all Transformer layers while still dedicating most of the silicon budget to the key MatMul kernels.
However, even when employing aggressive approximation, non-linearities in Transformers can become a very significant performance bottleneck when MatMul are hardware-accelerated.
%
\begin{figure}[tbp]
\centerline{\includegraphics[width=0.85\linewidth]{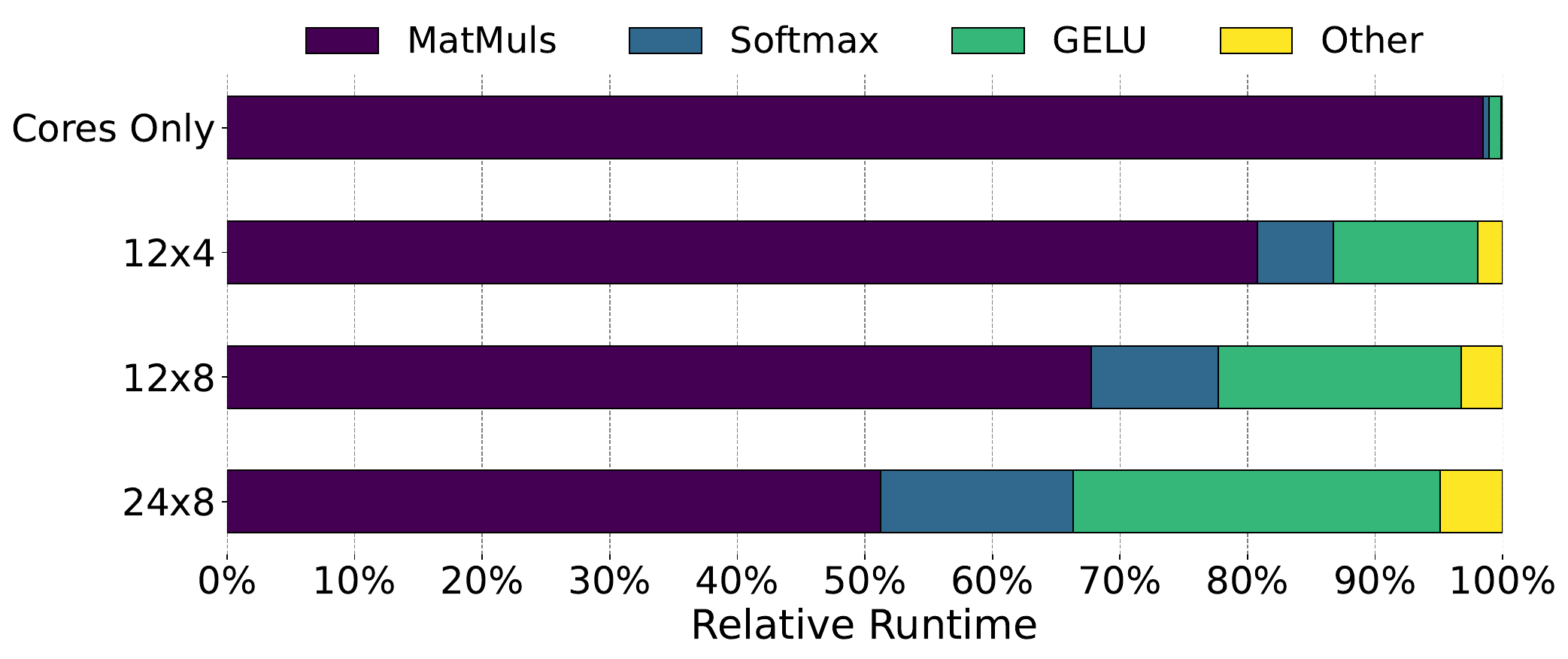}}
\caption{Breakdown of one of ViT's layers' runtime running on a 8 core PULP cluster enhanced with tensor processing units of various dimensions.}
\label{vit_scale}
\end{figure}
As a motivating example, we run an inference on ViT base~\cite{dosovitskiyImageWorth16x162020}, as an exemplary backbone for generative models such as BEiT~\cite{baoBEiTBERTPreTraining2022} and Florence-2~\cite{xiaoFlorence2AdvancingUnified2023}, on our baseline PULP cluster, integrating tensor processing units of various sizes.
For non-linearities, always executed on the RISC-V cores, we employ aggressive (and not accuracy-preserving) approximations using Schraudolph's method~\cite{schraudolphFastCompactApproximation1999} for exponentials and the sigmoid method for Gaussian Error Linear Unit (GELU)~\cite{hendrycksGaussianErrorLinear2023}.
As shown in Fig.~\ref{vit_scale}, the integration of a tensor processing unit with 12$\times$4 processing elements results in a 12.3$\times$ speedup compared to 8-core software.
However, increasing the tensor unit's size further does not yield a proportionally larger speedup due to the \textit{softmax} and \textit{GELU} non-linearities.
In fact, a 4$\times$ larger RedMulE instance results in only an additional 2.54$\times$ speedup, $\sim$63\% of the expected one.
The decreasing efficacy of hardware acceleration severely impacts energy efficiency. 

Unfortunately, non-linearities such as softmax and GELU cannot be easily replaced with other functions; in fact, they are essential to retain high accuracy in many GenAI transformers such as e.g. Whisper~\cite{radfordRobustSpeechRecognition2022} and DaViT~\cite{dingDaViTDualAttention2022}.
However, their highly non-linear nature makes it hard to implement them in hardware with simple circuits that retain high accuracy: in particular, the non-element-wise nature of softmax requires at least two passes on the input set and the usage of an exponential function, whereas GELU is based on the Gaussian cumulative distribution function, which can not be expressed in terms of elementary functions.

In this work, we propose multiple contributions to the State-of-the-Art to bypass the non-linearity bottleneck and enable the execution of perceptive and generative Transformers on edge devices, unleashing the full potential of embedded tensor processing units.
First, we introduce $\mathrm{expp}$, a fast, hardware-friendly approximation for floating-point $\exp$ by enhancing Schraudolph's method~\cite{schraudolphFastCompactApproximation1999} with a polynomial correction.
We exploit $\mathrm{expp}$ as an $\exp$ replacement both within softmax and within GELU.
Second, we propose \textit{SoftEx}, a parametric hardware accelerator for softmax and GELU non-linearities, based on our proposed fast approximate $\mathrm{expp}$ in BFloat16 (BF16) precision.
Finally, we introduce a flexible Transformer acceleration template that integrates SoftEx within an 8-core RISC-V PULP cluster along with a 24$\times$8 RedMulE tensor processing unit; we show how the introduction of SoftEx enables the cluster to achieve near-ideal speedup in the end-to-end execution of GenAI transformers inference.
Our contributions are as follows:
\begin{itemize}
    \item a hardware-friendly floating-point $\exp$ approximation, with 13$\times$ lower mean relative error and 3.7$\times$ lower maximum relative error compared to Schraudolph's method.
    \item SoftEx, a parametric hardware accelerator for softmax and GELU non-linearities. Compared to 8 RISC-V cores exploiting the less precise Schraudolph strategy, SoftEx cuts the total time spent on attention probabilities by up to 10.8$\times$ while requiring up to $26.8\times$ less energy; when computing GELU, using SoftEx speeds up the function by 5.11$\times$ and reduces its energy requirements by 5.29$\times$ with respect to the 8 cores exploiting the sigmoid approximation.
    \item a fully placed-and-routed heterogeneous acceleration cluster optimized for full-accuracy BF16 Transformers. In GlobalFoundries 12nm technology, the cluster achieves a peak performance of 310 GOPS at 0.80V or a peak efficiency of 1.34 TOPS/W at 0.55V on ViT base, including all overheads.
    \item a scalability analysis of the cluster when integrated in a large homogeneous compute mesh. In a $8\times 8$ mesh, on GPT-2 XL the single clusters achieve an average throughput of 285 GOPS, 82.6\% of that of a $1\times 1$ configuration, with an ensemble throughput of 18.2 TOPS.
\end{itemize}

The rest of this paper is organized as follows: Section~\ref{sec:related} discusses related work; Section~\ref{sec:background} introduces background material for our contribution; Section~\ref{expu} discusses the proposed exponential function approximation; Section~\ref{sec:architecture} introduces the architecture of SoftEx and of the Transformer acceleration cluster; Section~\ref{sec:accuracy} discusses the accuracy of the proposed approximations; Section~\ref{sec:implementation} presents post-layout results and performance evaluations; Section~\ref{sec:scalability} discusses the scalability of the proposed architecture; Section~\ref{sec:soa} compares our work with the State-of-the-Art. Finally, Section~\ref{sec:conclusion} concludes the paper.

\section{Related Work}
\label{sec:related}

To better position our work on perceptive and generative AI acceleration, relying on high-accuracy approximation of softmax and GELU, we analyze the State-of-the-Art in three aspects: techniques used to approximate the exponential function; accelerators and coprocessors dedicated to softmax and GELU; and finally complete architectures dedicated to the execution of Transformer-based AI.

\subsection{Approximations of the exponential function}
Exponentiation is historically a critical operation in neural networks, particularly in activation functions like softmax, where computational efficiency is crucial for real-time or large-scale applications. To address the high cost of precise exponentiation, several approximate algorithms have been developed. A very fast approximation commonly used in software implementations targeting high feed-forward speed is Schraudolph's method~\cite{schraudolphFastCompactApproximation1999}, which exploits the structure of floating point numbers to estimate the exponential with only a multiplication and an addition. This efficiency comes at the expense of accuracy, which Malossi~et~al.~\cite{malossiFastExponentialComputation2015} have addressed through a precision-adjustable polynomial correction of the mantissa. However, this refinement introduces additional floating-point operations, increasing computational overhead proportionally to the degree of the used polynomial.

In hardware accelerators, particularly those designed for fixed-point arithmetic and targeting low precision, it is a common practice to approximate the exponential function using lookup tables (LUTs)~\cite{sunHighSpeedSoftMax2018,gaoDesignImplementationApproximate2020,zhuEfficientPrecisionAdjustableArchitecture2020}.
Sun~et~al.~\cite{sunHighSpeedSoftMax2018} proposed a method that partitions the input into multiple components, computes their exponentials individually via LUTs, and then multiplies them to approximate the overall exponential.
Gao~et~al.~\cite{gaoDesignImplementationApproximate2020} reduced the reliance on LUTs by introducing an approach where LUTs are used for the high-order bits only, while a Taylor series expansion refines the computation for the least significant bits.
Zhu~et~el.~\cite{zhuEfficientPrecisionAdjustableArchitecture2020}, used LUTs to store the coefficients of a piecewise linear fitting of $\exp$ rather than its values.

Compared to Malossi~et~al. the method we propose is more efficient, as it applies the correction on the output mantissa only, enabling its calculation entirely in integer arithmetic. Moreover, compared to methods relying on LUTs, our approach is more hardware-friendly as it does not require storing any precomputed value in advance.

\subsection{Accelerators for softmax and GELU}
Even before the surge in popularity of attention-based networks, numerous studies have proposed optimizations to accelerate the softmax.
One common approach~\cite{yuanEfficientHardwareArchitecture2016,zhuEfficientPrecisionAdjustableArchitecture2020,tambe22912nm181TFLOPs2023,zhangBase2SoftmaxFunction2022} is to avoid explicit division by subtracting the logarithm of the denominator from each input. This approach effectively embeds the normalization step into the exponentiation process, simplifying the operation, requiring, however, specialized hardware to compute the logarithmic function.
Other works~\cite{keller17956TOPSDeep2022,zhangBase2SoftmaxFunction2022} have focused on simplifying the design by utilizing base-2 exponentials instead of the natural $\exp$. While this approach improves hardware efficiency, it requires fine-tuning of models to adapt to the altered activation function, limiting its applicability to pre-trained networks.
Additionally, both Keller~et~al.~\cite{keller17956TOPSDeep2022} and Wiese~et~al.~\cite{wieseAttentionbasedTinyMLHeterogeneous2024} enhance efficiency by skipping the initial maximum search. Instead, they compute the denominator online using a temporary maximum value, which is updated dynamically whenever a new maximum is encountered.
A more general method for approximating any nonlinearity within Transformer networks, introduced by Yu~et~al.~\cite{yuNNLUTNeuralApproximation2022} and applied in ViTA~\cite{chenViTAHighlyEfficient2024}, involves training a two-layer fully-connected neural network with ReLU activation to replicate the nonlinear functions. This network is then replaced by a look-up table, enabling the approximation of these functions through a single look-up operation and one multiply-accumulate. To enhance accuracy, Yu et al. further propose fine-tuning the two-layer network on the target model, which however limits the fine-tuned hardware to a single specific model.

In our proposed method, we require no quantization of the original network nor fine-tuning for the different bases of the exponentials. Moreover, our hardware is not tied to any particular model while retaining good performance and accuracy.



\subsection{Accelerators and SoCs for Transformers}
Transformer inference accelerators commonly target integer formats~\cite{wieseAttentionbasedTinyMLHeterogeneous2024,keller17956TOPSDeep2022,chenViTAHighlyEfficient2024,dumoulinEnablingEfficientHardware2024}, yielding high power and area efficiency compared to designs that use floating-point formats at the cost of the need of model quantization (including both weights and activations), which however is not always possible as models could be trained on large, non-publicly available datasets or with human feedback. Other processors, like the one developed by Tambe~et~al.~\cite{tambe22912nm181TFLOPs2023}, mix various techniques: very low-bitwidth floating-point (down to 8 or 4 bits); efficient handling of sparse matrices; and the usage of early exit algorithms, avoiding the computation of superfluous layers altogether. 
However, the applicability of all of these techniques is very limited when targeting Transformer networks that have not been retrained for this specific purpose.
ViTA~\cite{chenViTAHighlyEfficient2024}, achieves major area and efficiency gains by optimizing its architecture specifically for the Vision Transformer, significantly reducing memory usage and data movement. Similarly, Dumoulin et al.~\cite{dumoulinEnablingEfficientHardware2024} specifically focus on hybrid CNN-Transformers vision models, optimizing the execution of such architectures exploiting operation reordering and layer fusions, but limiting the accelerator to a single class of models.
The large scale of high-performance generative models has led to the development of large Neural Processing Units (NPU) to accelerate the massive amount of operations required by these models, such as SambaNova SN10 RDU~\cite{prabhakarSambaNovaSN10RDU2022} or Wormhole~\cite{ignjatovicWormholeAITraining2022}. However, the huge size and power consumption of these designs makes their deployment outside large data centers unfeasible.
A trend that is gaining momentum is to increase the efficiency of linear operations using block floating-point formats, which have proven to significantly increase throughput and energy efficiency in both inference~\cite{darvishrouhaniPushingLimitsNarrow2020} and training~\cite{nohFlexBlockFlexibleDNN2023}. Two examples of training accelerators exploiting these formats are Noh~et~al.~\cite{nohFlexBlockFlexibleDNN2023} and Zhang~et~al.~\cite{qianzhangFASTDNNTraining2022}.

In our work, we focused on a solution that combines good performance and energy efficiency with full flexibility and the capability to run unmodified networks without any quantization-aware fine-tuning. While block floating-point formats have proven highly effective for accelerating linear operations, this work focusing on addressing the significant bottlenecks posed by nonlinear operations, which remain a critical challenge in hardware acceleration of Transformers.

\section{Background}
\label{sec:background}

\subsection{Transformers}
Transformer networks consist of a sequence of encoder and/or decoder blocks, each containing an attention layer followed by a feed-forward neural network. 
In the \textit{attention} layer, the input of size $n\times d$, where $n$ is the sequence length and $d$ is the embedding size, is transformed through three linear projections into the Query ($Q$), Key ($K$) and Value ($V$) matrices, all of size $n\times d_h$. $Q$ and the $K^T$ are then multiplied and the softmax function is applied row-wise to the resulting $n\times n$ matrix to normalize each row to probabilities. Finally, the output of the softmax function is multiplied by $V$, obtaining an $n\times d_h$ matrix. This operation is performed $h$ times on different $Q$, $V$, and $K$ matrices, setting $d_h$ equal to $\frac{d}{h}$, to obtain multiple attention heads, which are concatenated and linearly transformed to produce the output of the attention layer, an  $n\times d$ matrix, the same size of the input.


In both encoders and decoders, the outputs of the attention layer are fed to a fully-connected feed-forward neural network. In the original Transformer model, this network applies two linear transformations to the inputs with a ReLU nonlinearity in the middle. Later Transformer-based models have commonly increased the number of linear transformations (e.g., MobileBERT \cite{sunMobileBERTCompactTaskAgnostic2020}) or replaced ReLU with other activation functions, in particular with GELU (e.g., ViT~\cite{dosovitskiyImageWorth16x162020}, GPT-2~\cite{radfordLanguageModelsAre2019}).

\subsection{Softmax}
Softmax is an extremely common activation function found in many neural networks and essential to the multi-head self-attention mechanism used in Transformers.
Its purpose is to normalize a vector of real-valued \textit{scores} to a vector of probabilities that sum up to 1; it acts like a softening of the $\mathrm{argmax}(\cdot)$ function. Given a vector $x$ of length $N$, the softmax function is defined as:
\begin{equation}
    \label{eq:softmax}
    \mathrm{Softmax}(x)_i=\frac{e^{x_i-\max(x)}}{\sum_{j=1}^{N}{e^{x_j-\max(x)}}}
\end{equation}
Subtracting the maximum is necessary to ensure numerical stability, as it guarantees that all inputs of the exponential will be $\leq 0$.
Calculating this function poses two main challenges: one is the fact that it is a function based on the natural exponential, a transcendental function, and the other is that softmax is not a pointwise function, as it requires both a maximum search and a summation over every element. 

The formulation of the function shown in Eq. \ref{eq:softmax} requires four memory accesses per element: one read for the initial maximum search, another for the calculation of the denominator, and one read plus one write for the normalization. 
It is possible to skip the initial maximum search, thus saving a vector read, by employing an online normalization scheme: instead of searching for the global maximum before starting the calculation of the denominator, the two steps are fused into a single pass where each score is subtracted the current local maximum and the denominator is dynamically updated whenever a new maximum value is encountered. 

Let $\mathrm{Den}(x,n)$ be the value of the denominator of the Softmax function of the first $n$ elements of a vector $x$, and $\mathrm{Max}(x,n)$ the maximum value among the first $n$ elements of a vector $x$, for a generic positive integer $N$ we have: 
\begin{equation}
    \begin{aligned}
        \mathrm{Den}(x, N) &= \sum_{i=1}^{N} e^{x_i - \mathrm{Max}(x,N)} \\
                           &= \mathrm{Den}(x, N-1) \cdot e^{\mathrm{Max}(x,N-1) - \mathrm{Max}(x,N)} \\
                           &\quad + e^{x_N - \mathrm{Max}(x,N)}
    \end{aligned}
\end{equation}
Therefore, whenever the maximum score is updated, the current partial denominator has to be multiplied by the exponential of the difference between the current maximum and the new maximum before adding the new exponentiated score.

\subsection{GELU}\label{bg_gelu}
The GELU activation function is a widely adopted nonlinearity in deep learning due to its superior performances compared to other activation functions such as ReLU, and is found in multiple tranformer-based models such as BERT \cite{sunMobileBERTCompactTaskAgnostic2020} and the Vision Transformer \cite{dosovitskiyImageWorth16x162020}. The main idea behind this function is to build a smooth and differentiable approximation of the ReLU activation gradually shifting the input weights from zero to one, thus avoiding the singularity in $x = 0$ and adding a curvature for every $x$.
In particular, Hendrycks and Gimpel propose to weight the inputs by the value of the Gaussian cumulative distribution function (CDF), $\Phi(\cdot)$:
\begin{equation}\label{gelu_base}
    \mathrm{GELU}(x) = x \cdot \Phi(x) = x \cdot \frac{1}{\sqrt{2\pi}}\int_{-\infty}^{x}{\exp \left(- \frac{t^2}{2} \right) dt}
\end{equation} 

However, implementing the Gaussian CDF accurately is computationally intensive and impractical for many real-world applications. As an example, the default exact implementation used in the PyTorch Python library is based on Eq.~\ref{gelu_base}, with $\Phi$ implemented as a piecewise seventh-order polynomial\footnote{\url{https://github.com/bminor/glibc/blob/master/sysdeps/ieee754/dbl-64/s_erf.c\#L199}}.

To address the high complexity of the exact implementations of $\Phi$, the original authors of the GELU activation function introduced two approximations that significantly ease implementation. The first, a more precise approximation, utilizes the hyperbolic tangent function:
\begin{equation}\label{gelu_tanh}
    \mathrm{GELU}(x) \approx \frac{1}{2}x\left(1 + \tanh\left(\sqrt{\frac{2}{\pi}}\left(x + \frac{11}{123} x^3\right)\right)\right)
\end{equation}
while the second, simpler approximation relies on the sigmoid function:
\begin{equation}\label{gelu_sigmoid}
    \mathrm{GELU(x)} \approx x\sigma(1.702 x)
\end{equation}

Despite their efficiency compared to exact implementations, these approximations are still challenging to implement in low-power, high-throughput systems, as both the hyperbolic tangent and sigmoid functions involve division.

In this work, we calculate the $\Phi$ function using an approximation derived from work by Chiani~et~al.~\cite{chianiNewExponentialBounds2003} and Tanash~and~Riihonen~\cite{tanashGlobalMinimaxApproximations2020}.
Chiani et al.~\cite{chianiNewExponentialBounds2003} observe that for non-negative $x$, $1-\Phi(x)$ can be bound by a sum of exponentials:
\begin{equation}\label{q_chiani}
    \begin{aligned}
        1-\Phi(x) &\leq \sum_{i=1}^N a_i e^{-b_i x^2}, x\geq0
    \end{aligned}
\end{equation}
Building on this result, Tanash and Riihonen~\cite{tanashGlobalMinimaxApproximations2020} propose a method to optimize the $a$ and $b$ parameters by solving a system of equations with five unknowns --  the optimal values of $a$ and $b$, the $2N$ maximum error points $x_k$, and the maximum relative error $r_{\mathrm{max}}$:
\begin{equation}\label{q_}
    \begin{aligned}
        \begin{cases}
            r'(x_k) = 0, & \mathrm{for} \; k = 1,2,...,2N, \\
            r(x_k) = (-1)^{k+1}r_\mathrm{max}, & \mathrm{for} \; k = 1,2,...,2N, \\
                \sum_{n=1}^{N}a_n = \frac{1}{2}, & \mathrm{when} \; r(0) = 0, \\
                \sum_{n=1}^{N}a_n = \frac{1}{2} - \frac{r_{\mathrm{max}}}{2}, & \mathrm{when} \; r(0) = -r_{\mathrm{max}} \\
            r(x_{2N+1}) = -r_{\mathrm{max}}.
        \end{cases}
    \end{aligned}
\end{equation}
We defer to Appendix I for a full treatment of this approximation.
Using this sum-of-exponentials formulation, we approximate the Gaussian CDF with an adjustable precision and avoid expensive dividers. 
As this formulation is symmetric, for $x < 0$, Eq.~\ref{q_} computes $\Phi(x)$ directly rather than $1-\Phi(x)$.
The GELU function can thus be implemented following Algorithm~\ref{gelu_sum}.
\RestyleAlgo{ruled}
\begin{algorithm}[tb]
\footnotesize
\caption{GELU calculation using a sum of $\exp$.}
\label{gelu_sum}
\KwData{$x$, $a = [a_1, a_2, \dots, a_N]$, $b = [b_1, b_2, \dots, b_N]$}
\KwResult{$y = \mathrm{GELU}(x)$}
$x_{sq} \gets x^2$\;
$s \gets 0$\;
\For{$i \gets 1$ \KwTo $N$}{
    $s \gets s + b_i \cdot \exp(-a_i x_{sq})$\;
}
\uIf{$x \geq 0$}{
    $y \gets x\cdot(1 - s)$\;
}
\Else{
    $y \gets x\cdot s$\;
}
\Return $y$\;
\end{algorithm}

\section{Exponential Function Approximation}\label{expu}
\begin{figure}[tbp]
\centerline{\includegraphics[width=0.85\linewidth]{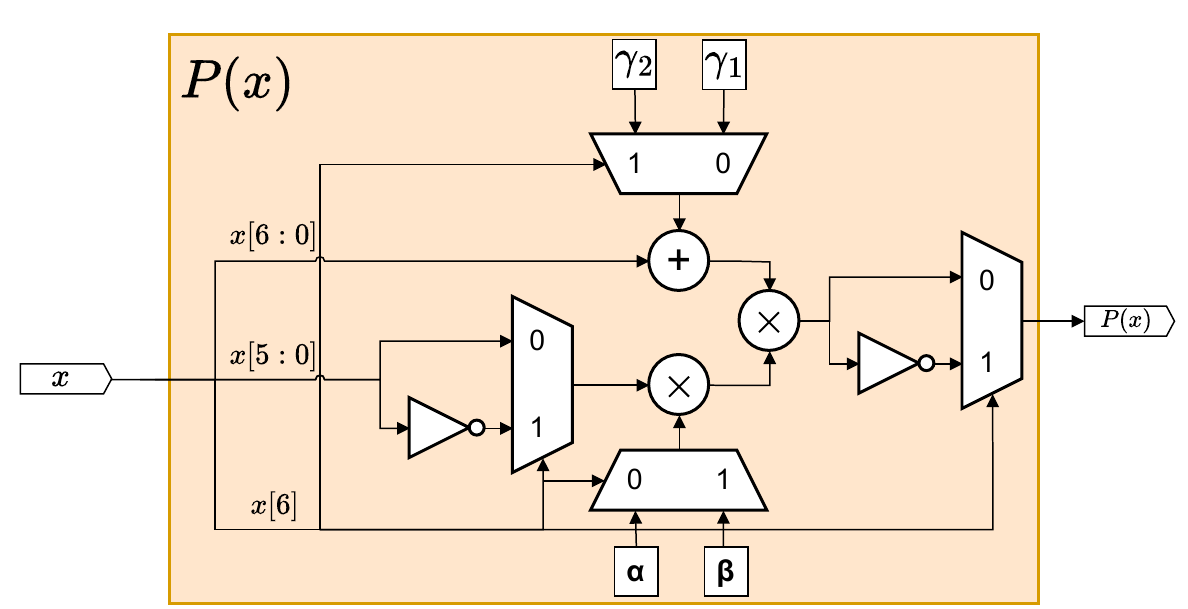}}
\caption{The circuit implementing the correction proposed in Section \ref{expu}, assuming a 7-bit mantissa in BFloat16.}
\label{mant_correction}
\end{figure}

Our first contribution is a novel exponentiation algorithm based on an extension of Schraudolph's method~\cite{schraudolphFastCompactApproximation1999} applied to BF16 inputs in conjunction with a polynomial correction of the mantissa, whose implementation is depicted in Fig.~\ref{mant_correction}.
%
%
%
%
Schraudolph's method, which we will refer to as $\mathrm{exps}(\cdot)$, is based on the following approximation: we move to base 2 to simplify the exponentiation, defining $x'=x\cdot\frac{1}{\log 2}$; then,
\begin{equation}
    \exp (x) = 2^{x'} \approx
        2^{\mathrm{int}(x')} \cdot \big(1+\mathrm{frac}(x')\big) \doteq
        \mathrm{exps} (x)
    \label{eq:schraudolph}
\end{equation}
It can be shown that this approximation can be implemented as a three-step procedure on a BF16 input $x$, described in Algorithm~\ref{sch_alg}.
\RestyleAlgo{ruled}
\begin{algorithm}[tb]
\footnotesize
\caption{Schraudolph's method on BF16 inputs.}
\label{sch_alg}
\KwData{$x$, a BF16 number}
\KwResult{$y = \mathrm{exps}(x)$}
$m \gets \mathrm{mantissa}(x)$\;
$e \gets \mathrm{exponent}(x)$\;
$s \gets \mathrm{sign}(x)$\;
$b \gets \mathrm{bias}(x)$\;

\If{$s = 1$}{
$m \gets -m$\;
}

$scale \gets \frac{1}{\log2} \ll{(e - b + 7)}$\;
$b_{sh} \gets b \ll{7}$\;

$m_{sh} \gets m \cdot scale + b_{sh}$\;

\uIf{$\mathrm{abs}(m_{sh}) \geq 2^{16}$}{
    \uIf{s = 0}{
        $y \gets +\infty$\;
    }
    \Else{
        $y \gets 0$\;
    }
}
\Else{
    $y \gets \mathrm{reinterpretAsFloat}\left(m_{sh}\right)$\;
}
\Return $y$\;
\end{algorithm}

For details and proof, we defer to the original paper~\cite{schraudolphFastCompactApproximation1999} or our open-source implementation\footnote{\url{https://github.com/belanoa/softex/tree/gelu/rtl}}.

%
To improve the mantissa's accuracy, we propose to replace the $\big(1+\mathrm{frac}(x')\big)$ factor in Eq.~\ref{eq:schraudolph}, which approximates the precise value $2^{\mathrm{frac}(x')}$, with a polynomial $\big(1+P(\mathrm{frac}(x'))\big)$.
We split the $[0,1)$ range in two halves such that we can use the mantissa's most significant bit to discriminate them, using two distinct second-order polynomials in a form similar to $1+ax(x+b)$ to approximate $2^{\mathrm{frac}(x')}$.
We choose this particular form as it eases hardware implementation.
Our approximation is thus
\begin{equation}
    \mathrm{expp}(x) \doteq 2^{\mathrm{int}(x')} \cdot \big(1+P(\mathrm{frac}(x'))\big)
\end{equation}
We define $P(\cdot)$ as follows. We start with an intuitive approximation: if the input $x$ falls into the $[0, 0.5)$ range, i.e., the most significant bit is 0, we approximate $2^x-1$ with the sum between a tangent straight line in $x=0$ and a parabola centered in $x=0$:
\begin{equation}\label{exp_half_1}
P(x) = \log 2\cdot x+\alpha x^2,\; x\in[0,0.5)
\end{equation}
This function can be rewritten in the form $ax(x+b)$ by applying the following simple transformation:
 \begin{equation}\label{exp_half_1_final}
  P(x) = \log 2\cdot x+\alpha x^2
  = \alpha x \left(x+\frac{\log 2}{\alpha}\right),\; x\in[0,0.5)
 \end{equation}
If $x$ falls into the $[0.5, 1)$ range, i.e., the most significant bit is 1, we use the sum between a straight line tangent to $2^{x}-1$ in $x=1$ and a parabola centered in $x=1$:
\begin{equation}\label{exp_half_2}
    P(x) = 2\log{2}\cdot x+1-2\log{2} + \beta(1-x)^2, \; x\in[0.5,1)
\end{equation}
To reduce Eq.~\ref{exp_half_2} to the form $ax(x+b)$, we note that $1-x$ can be further approximated as the one's complement of $x$, i.e., $\mathrm{not}(x)$:\\
\begin{equation}\label{exp_half_2_final}
    \begin{aligned}
        P(x) &=         2\log{2}\cdot x+1-2\log{2} + \beta(1-x)^2                               \\
             &=         1 - 2\log{2}\cdot(1-x) + \beta(1-x)^2                                   \\
             &=         1 - \beta(1-x)\left(x + \frac{2\log{2}}{\beta} - 1\right)               \\
             &\approx   \mathrm{not}\left(\beta\,\mathrm{not}({x}) \cdot\left(x+\frac{2\log{2}}{\beta}-1\right)\right), \; x\in[0.5,1)
    \end{aligned}
\end{equation}

In practice, we replace the additive factors in Eqs.~\ref{exp_half_1_final}, \ref{exp_half_2_final} with two free parameters $\gamma_1$, $\gamma_2$. The final expression of $P(\cdot)$ is therefore
\begin{align}
    P(x) &\doteq \alpha x \left(x+\gamma_1\right),\; &x\in[0,0.5) \\
    P(x) &\doteq \mathrm{not}\big(\beta\,\mathrm{not}({x}) \cdot\left(x+\gamma_2\right)\big), \; &x\in[0.5,1)
\end{align}

We derive parameters that minimize the error introduced in $\mathrm{expp}$ compared to $\mathrm{exp}$ by using a heuristic Monte Carlo procedure with $10^6$ trials.
The final values we used for the rest of this work are $\alpha=0.21875$, $\beta=0.4375$, $\gamma_1=3.296875$, and $\gamma_2=2.171875$. To represent the $\alpha$ and $\beta$ parameters we use a 4-bit integer number with a fixed scaling factor of $2^{-4}$ while $\gamma_1$ and $\gamma_2$ are represented using an 8-bit integer with a scaling factor of $2^{-6}$. 


\section{Architecture}
\label{sec:architecture}

In this Section, we discuss the architecture of our template for perceptive and generative AI acceleration.
First, we review the baseline architecture of our template, based on a heterogeneous Parallel Ultra-Low Power (PULP) cluster.
Then, we focus in detail on the microarchitecture of SoftEx, which is key to enable our template to achieve effective acceleration in end-to-end Transformer execution.

\begin{figure}[tbp]
\centerline{
\includegraphics[width=0.85\columnwidth]{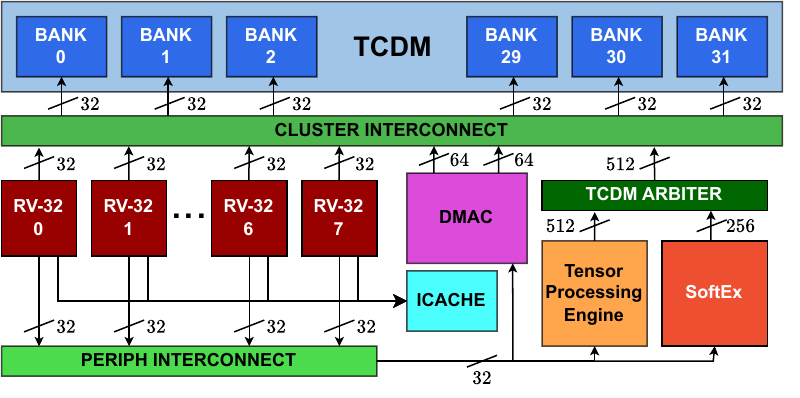}
}
\caption{Architecture of the enhanced PULP cluster proposed in this work. External connections are not shown for simplicity.}
\label{cluster}
\end{figure}

\begin{figure*}[tbp]
\centerline{
\includegraphics[width=1.85\columnwidth]{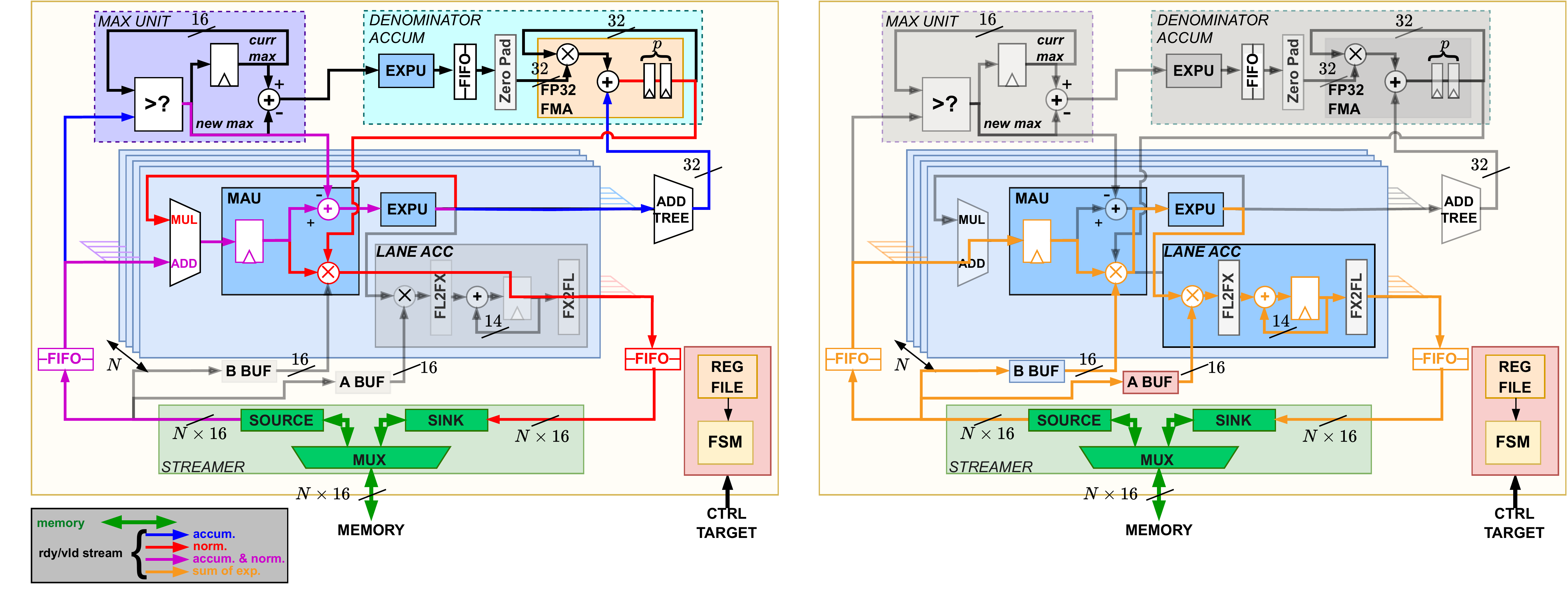}
}
\caption{A detailed view of SoftEx and its Datapath. In the left image, the paths used in the calculation of Softmax are highlighted, with paths used in the accumulation step highlighted in blue, those used in the normalization step highlighted in red, and those used in both steps highlighted in purple. In the right image, paths used in the sum of exponentials calculation are highlighted in orange. Unused paths in a mode are grayed out.}\label{datapath}
\end{figure*}

\subsection{The PULP Cluster}
Our work relies on a compute cluster based on the PULP (Parallel Ultra-Low Power) parametric heterogeneous architecture template.
The baseline PULP cluster features 2--8 32-bit RISC-V digital signal processing cores accessing a shared scratchpad memory, called Tightly-Coupled Data Memory (TCDM), and a shared 32KiB instruction cache. To accelerate specific tasks, the cluster can employ instruction set extensions in the RISC-V cores and/or cooperative HWPEs, which share the TCDM with the cores at L1.
In this work, we focus on a cluster configuration featuring 8 RISC-V \texttt{RV32IMFCXpulpnn} cores and 256KiB of TCDM split among 32 banks. The cores have private FPUs supporting FP32 and BF16 formats. Moreover, to accelerate the typical workloads in Transformers, we integrate a large tensor processing unit with 24$\times$8 BF16 FMAs based on the RedMulE architecture~\cite{tortorellaRedMuleMixedprecisionMatrix2023} together with an instance of SoftEx to enable the acceleration of typical nonlinearities found in Transformers. 
Fig.~\ref{cluster} shows the detailed architecture of the target accelerator cluster.

\subsection{SoftEx softmax and GELU accelerator}
\subsubsection{Architectural overview}
SoftEx is parametric hardware accelerator for the softmax and GELU functions targeting BF16 vectors. SoftEx is implemented in SystemVerilog HDL and is organized as a Hardware Processing Engine (HWPE)\footnote{\url{https://hwpe-doc.readthedocs.io}}, as shown in Fig.~\ref{datapath}, meant for integration in a System-on-Chip as a cooperative unit; it comprises a programming interface with an internal control finite-state machine (\textit{controller}), a set of specialized data movers to move data to/from memory (\textit{streamer}), and a specialized datapath.

The datapath features a configurable number of lanes ($N$), each comprising a BF16 Multiplication and Addition Unit (MAU), a BF16 Exponential Unit (EXPU) implementing the algorithm described in Section~\ref{expu}, and a Lane Accumulator.
SoftEx also contains an Accumulator module with a single pipelined FP32 Fused Multiply-Add (FMA) unit, used during the computation of softmax to calculate first the value of the denominator and then its reciprocal, used to normalize the output as presented in Eq.~\ref{eq:softmax}.
We perform the accumulation in higher precision than the rest of the datapath because the contributions from relatively small inputs, generally the majority, would otherwise be lost.

SoftEx is designed to fetch data from a tightly-coupled memory which could be subject to conflicts on the banks,
hence external communication happens by means of a request/grant memory protocol (shown in Fig.~\ref{datapath} as double-ended arrows), whereas internal communications in the datapath (shown as single-ended arrows) employ a ready/valid handshake scheme, to enable correct backpressure support both from memory access bubbles and from internal operations.
%

\subsubsection{Calculation of softmax}
The calculation of the softmax function is divided into three steps: \textit{accumulation}, where the maximum score is found and the denominator of Eq.~\ref{eq:softmax} is calculated; \textit{inversion}, where the reciprocal of the denominator is computed; and \textit{normalization}, where each of the output probabilities is calculated.

\paragraph{Accumulation step}
During the \textit{accumulation} step, the accelerator reads $N$ inputs per cycle; the first array of BF16 MAUs subtracts the maximum score from every input, then the results are fed to the EXPUs. Finally, the exponentiated scores are added together and pushed into the accumulator.
We keep track of the current partial maximum in a dedicated buffer and use it to offset the scores, as shown in the MAU in Fig.~\ref{datapath}.
If a new input is larger than the current maximum, we use $\mathrm{e}^{\mathrm{currmax}-\mathrm{newmax}}$ to update the value of the denominator in the \textit{denominator accumulator}, before adding the new batch of exponentiated scores to the cumulated value.

More in detail, to avoid cumulating values with different scaling factors, each input to the denominator accumulator FMA is assigned a monotonically increasing tag.
When a new maximum is detected in the max unit, new operations to be inserted in the pipeline are stalled by lowering the denominator accumulator's FIFO ready signal, while in-flight operations are rescaled sequentially by $\mathrm{e}^{\mathrm{currmax}-\mathrm{newmax}}$ using the FMA itself according to their tag.
The mechanism supports correct accumulation even in the pathologic case of a monotonically increasing input.
The final result of the accumulation step is $\sum_j{\mathrm{expp}{(x_j-\max(x))}}$, stored directly in the last FMA pipeline stage.

\paragraph{Inversion step}
Once the computation of the denominator is complete, its reciprocal is computed using the Newton-Raphson method. As starting point for the Newton iteration, we note that given the denominator value is $(1+M)2^{E-B}$ (always non-negative), the exponent of its reciprocal can be computed exactly as $2B - 1 - E$; we estimate the mantissa with the parabola $\frac{(1 - M) ^ 2}{2}$, where $1-M$ is approximated with $\mathrm{not}(M)$. Starting from this initial value, the denominator accumulator's FMA is used to perform two Newton iterations.
At the end of this step, the $1/\sum_j{\mathrm{expp}{(x_j-\max(x))}}$ factor that is required for the following normalization step is statically stored in the denominator accumulator.

\paragraph{Normalization step}
Once the reciprocal of the denominator has been calculated, the normalization step begins. In this phase, the exponentiated values are multiplied by the reciprocal of the denominator (cast back to BF16) and written back to memory. The array of MAUs is used to both subtract the maximum score and multiply by the inverted denominator.
To fully utilize the available memory bandwidth during both the accumulation and normalization steps, the load of a new vector of scores and the store of output probabilities are alternated by exploiting the multiplexer shown in the streamer unit.
Since all streams in the datapath employ a ready/valid handshake scheme, this alternating pattern propagates to the MAUs without additional control overhead.

\subsubsection{Calculation of GELU} To compute the GELU activation function, we follow Algorithm~\ref{gelu_sum}. However, rather than accelerating the whole algorithm, we use SoftEx to accelerate the second and most computationally intensive step only, the sum of exponentials, while delegating the remaining, simpler steps to the cores.

In this step, each of the $N$ MAUs multiplies its respective input, by the current $b_i$ weight and feeds it to the exponential units. Then, the exponentiated results are pushed into the lane accumulators, which weight the inputs by the current $a_i$ weight and accumulate the results. Let $N_w$ the number of terms in the sum of exponentials, the inputs are held steady for $N_w$ cycles while the weights are updated every cycle.

Unlike the denominator accumulator, the lane accumulators adopt a simpler architecture, not based on FMAs. While the weighting of the inputs is still performed using floating-point multipliers, we note that, as the addends are all positive and Eq. \ref{q_} evaluates $1 - \Phi(x)$ for $x\geq0$ only, the accumulated value is bounded within the $(0,0.5]$ range, allowing us to perform the accumulation in fixed-point format rather than floating-point format, thus avoiding the complexity of floating-point additions. While this approach has the drawback of quantizing relatively small values to zero, we will show that in practice, if the internal accumulator is adequately sized, this effect has negligible impact on network accuracy. We implement the accumulator using a 14-bit fixed-point representation.

Once all $N_w$ weights have been processed, the accumulated values are converted back to BF16 format and pushed into the output FIFO, while a new vector of $N$ inputs is read, resulting in an output bandwidth of $\frac{N}{N_w}$ elements per cycle. 
To optimize memory accesses, when the $a$ and $b$ weight buffers are exhausted, the buffers are read in reverse order, eliminating the need to read a new set of weights at the conclusion of each sum of exponentials computation, thus enhancing efficiency.

\section{Accuracy Analysis}
\label{sec:accuracy}
\subsection{Exponentiation Algorithm} \label{exp_alg}
\subsubsection{Accuracy of expp}
We validated our $\mathrm{expp}$ approximation of the exponential function on $10^8$ samples from a uniform distribution in the $[-88.7,88.7]$ range, i.e., the range of inputs that do not cause any overflow when exponentiated in BF16 or FP32.
The results were compared with those obtained using Schraudolph's method ($\mathrm{exps}$) and using glibc's implementation as the baseline. With respect to glibc's implementation, the proposed exponentiation algorithm achieves a mean relative error of 0.14\% and a maximum relative error of 0.78\%. Compared to Schraudolph's method, our approximation has a $13\times$ lower mean relative error and a $3.7\times$ lower maximum relative error. Using a second-order Chebyshev polynomial, the algorithm devised by Malossi et al.~\cite{malossiFastExponentialComputation2015} achieves an average relative error of 0.11\%, 0.03\% lower than out algorithm, which however comes at the cost of two additional expensive floating-point FMAs, as the correction is applied directly on the input. 

\subsubsection{Accuracy of softmax}
We tested the effect of our approximation on the accuracy of the softmax function on vectors of 1024 elements from one of MobileBERT's~\cite{sunMobileBERTCompactTaskAgnostic2020} attention layers to simulate a real use case. 
The resulting mean relative error of the outputs is 0.44\%, $3.2\times$ lower than the same function calculated using Schraudolph's method as exponentiation algorithm.

To further evaluate the accuracy of the proposed approximation, we evaluated the effects of replacing all $\exp$ with $\mathrm{expp}$ on the final output of MobileBERT, employing the SQuAD v2 dataset~\cite{rajpurkarSQuAD100000Questions2016} and the CoLA corpus~\cite{warstadtNeuralNetworkAcceptability2019}.
We measured the deviation from the output logits of the base model calculated with an accurate $\exp$, versus using either Schraudolph's method ($\mathrm{exps}$) or our proposed $\mathrm{expp}$.
On SQuAD, the resulting mean squared error of the output logits when using $\mathrm{expp}$ is 0.0292, a 17.5\% reduction compared to Schraudolph's method. On CoLA the pattern is similar, with a $22.8\%$ reduction in the mean squared error compared to $\mathrm{exps}$ (from 0.0149 to 0.0115).

In conclusion, the approach we propose offers significant accuracy improvements over Schraudolph's method, both when used standalone and when used as part of more complex functions.
This justifies using the proposed approximation when targeting the execution of unmodified pretrained Transformers.

\begin{figure*}[tbp]
\centerline{
\includegraphics[width=1.8\columnwidth]{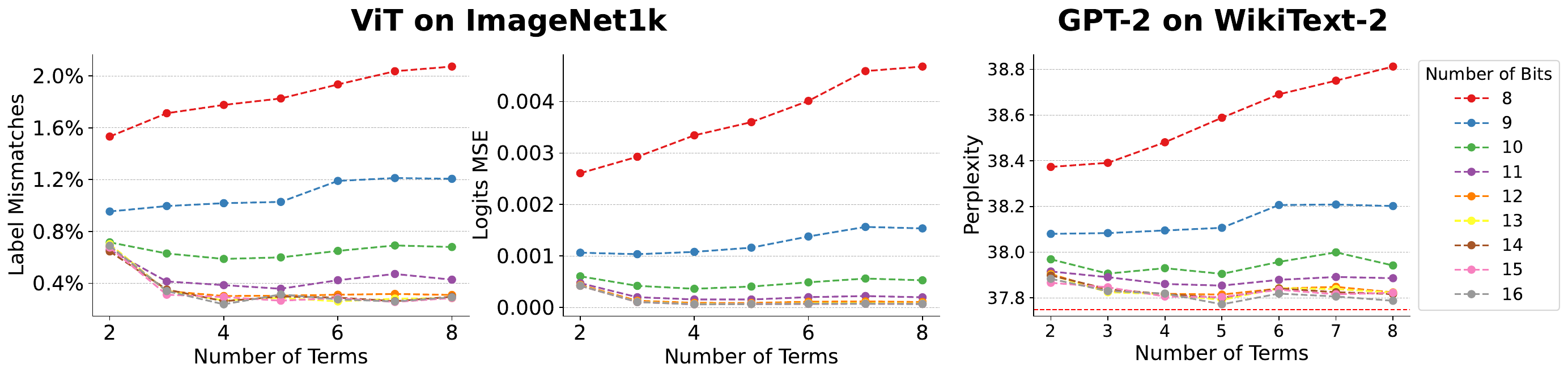}
}
\caption{The effects of changing the number of bits in the lane accumulators and number of terms in the sum of exponentials. From left to right: the number of mismatches in the predicted labels and the mean squared error (MSE) of the output logits of ViT on ImageNet1k, and the perplexity of GPT-2 on the WikiText-2 dataset. For ViT, the number of mismatches and the MSE are defined with respect to a model in which both the exponential function and GELU are computed using accurate methods. The dashed red line on the rightmost plot represents the perplexity of GPT-2 when both the exponential function and GELU are calculated with accurate methods.}
\label{gelu_acc}
\end{figure*}

\subsection{GELU Approximation} \label{gelu_alg}
We solved Eq. \ref{q_} for $r(0)=-r_{\mathrm{max}}$ and $x_{2N+1}=2.8$.
We chose these parameters as: \textit{1)} in the calculation of GELU, for values around 0, $\Phi$ is multiplied by near-zero values so we can safely decrease the maximum error of the approximation by making $x = 0$ a maximum error point; \textit{2)} for $x > 2.8$ the GELU function is equal to the input and for $x < -2.8$ the value of $\Phi$ rapidly approaches zero.

To assess the optimal number of terms in the sum-of-exponentials and bits in the accumulators, and evaluate the practical impact of our approximation on model performance, we validated the algorithm by testing it on ViT and GPT-2. In particular, we measured the deviation in predicted labels for the Vision Transformer on the ImageNet1k \cite{dengImageNetLargescaleHierarchical2009} validation set and the changes in perplexity for GPT-2 on the WikiText-2 \cite{merityPointerSentinelMixture2016} corpus. The results of this experiment are shown in Fig.~\ref{gelu_acc}.

On ImageNet1k and WikiText-2, using less than 10 bits for accumulators results in significant deviation from the base model.
With 11 or more bits, deviations stabilize with 4 terms used, reducing mismatches and mean squared error.
In WikiText-2, 8-9 bit accumulators lead to higher perplexity, but increasing terms reduces it. 
For configurations using 11 or more bits, increasing the number of terms generally reduces perplexity, with the optimal performance typically observed at 5 terms in WikiText-2 and 4 in ImageNet1k.
The accuracy degradation with accumulators using $\leq$10 bits and many terms is due to the fact that smaller addends in the sum of exponentials are truncated during the fixed-point conversion.
In both benchmarks, using four terms and 14 bits in the lane accumulator proves to be a good compromise between model accuracy and efficiency.

As a comparison, the software implementation with Schraudolph's method  and using the sigmoid approximation of GELU achieves a 4.96\% label mismatches and a 0.652 logits mean squared error on ImageNet1k, and a perplexity of 37.9243 on the WikiText-2 benchmark. If, instead of the sigmoid approximation, Eq.~\ref{gelu_tanh} is used to approximate GELU, the percentage of misamtches and the mean squared error decrease to 0.65\% and 0.0005 respectively, while perplexity decreases to 37.74.
With the proposed method, at four terms and 14 bits we achieve a label mismatch of 0.27\% and a mean squared error of $6.4\cdot10^{-5}$ on ImageNet1k and a perplexity of 37.816 on WikiText-2.

In conclusion, our method provides a precision-adjustable approach for approximating the GELU activation, with negligible effects on network performance when 4–5 terms and 14-bit accumulators are used.

\section{Experimental Results}
\label{sec:implementation}
\subsection{Experimental Setup}
Our experiments focus on SoftEx with $N=16$ lanes, resulting in a 256-bit memory interface.
The algorithms have been validated by implementing them in Python using the PyTorch library. The SoftEx-augmented cluster has been implemented in GlobalFoundries 12LP+ technology, targeting 700MHz frequency in worst-case conditions (SS process, $0.72$V and $125^{\circ}$C) using Synopsys Design Compiler for synthesis and Cadence Innovus for placement, clock tree synthesis, and routing. The system's power consumption has been extracted under typical conditions (TT process and $25^{\circ}$C) using Synopsys Primetime with annotated switching activity from a post-layout simulation in two operating points: at $0.80$V and 1.12 GHz to maximize throughput, and at $0.55$V and 460 MHz to maximize energy efficiency. 
Where not otherwise specified, we consider 1 MAC = 2 OPs.



\begin{figure}[tbp]
\centerline{\includegraphics[width=0.9\linewidth]{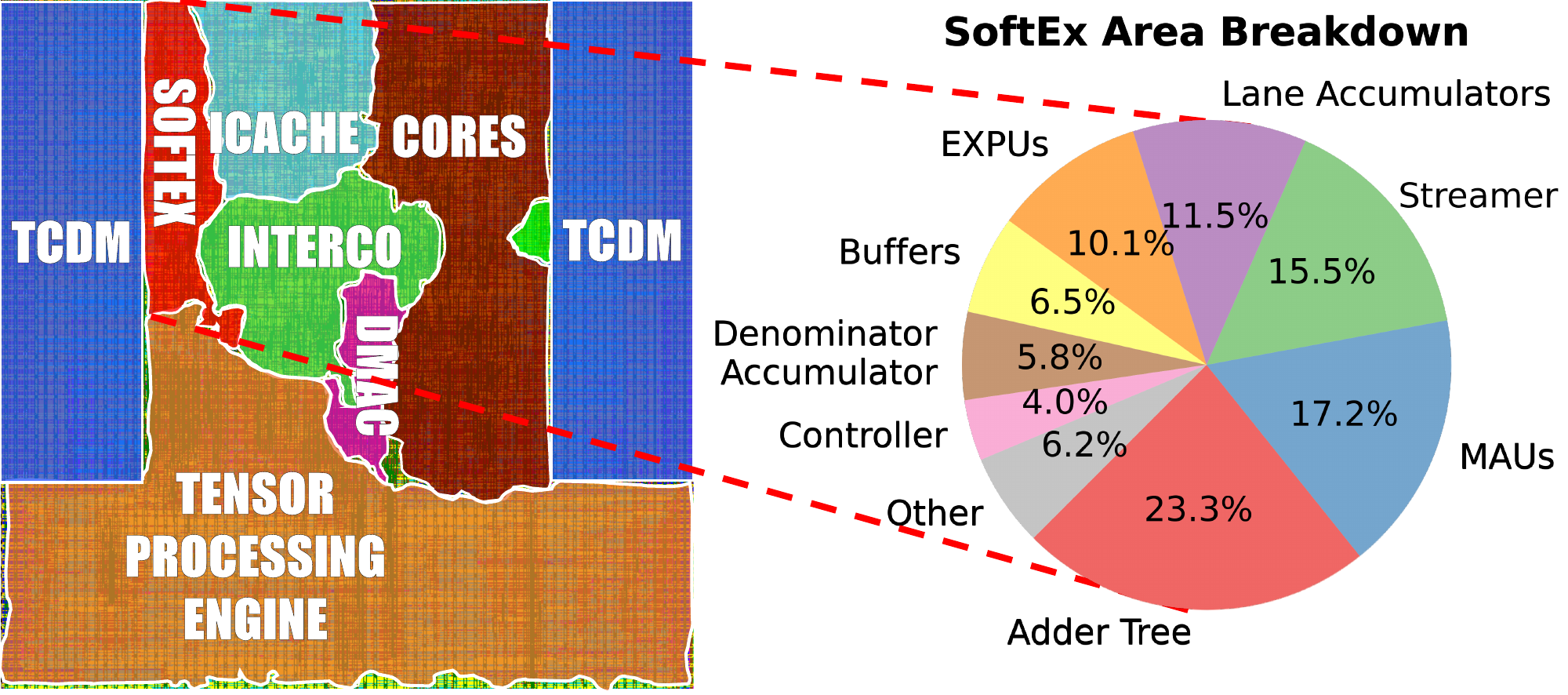}}
\caption{SoftEx's area breakdown and the annotated layout of the proposed cluster (1.1mm$\times$1.1mm).}
\label{breakdown}
\end{figure}

\subsection{SoftEx Area, Power and Performance}

\paragraph{Area breakdown}
Fig.~\ref{breakdown} shows the layout of the cluster and an area breakdown of the SoftEx accelerator.
SoftEx measures 0.039 mm$^2$, only 3.22\% of the cluster, which occupies 1.21 mm$^2$.  Due to its higher precision, the adder tree is the largest contributor to the area of the accelerator, accounting for 23.3\% of the total, followed by the MAUs (17.2\%) and the Streamer (15.5\%). With an overall contribution of 11.5\% and 10.1\%, the lane accumulators and the exponential units are only the fourth and fifth largest contributors respectively, validating our choice to perform the sum of exponentials in fixed-point format and to use a slightly more complex datapath to implement $\mathrm{expp}$ instead of Schraudolph's method.

\paragraph{Power breakdown}
When performing softmax, the average power consumption of the cluster is 278 mW (53.2 mW for SoftEx, with the rest dominated by the TCDM's SRAM banks) at 0.80V or 56.1 mW at 0.55V (9.87 mW for SoftEx).
Inside the accelerator, the MAUs dominate the power consumption with a contribution of 24.2\%, while the adder tree, although much larger than the MAUs, accounts on average for 10.5\% of the total because it is only used during the accumulation step. The exponential units contribute only 13.7\% of SoftEx's total power.

When performing the sum of exponentials in the GELU calculation, the average power consumption of the cluster is is 276 mW at 0.80V (50.8 mW for SoftEx) or 55.7 mW at 0.55V (9.46 mW for SoftEx). With a contribution of 22\%, the lane accumulators dominate SoftEx's power consumption, with the MAUs following closely at 20\% of the total. Due to their higher utilization compared to the calculation of softmax, the exponential units contribution is larger in a sum of exponentials compared to softmax, accounting for 16\% of the total.  

\begin{figure}[tbp]
\centerline{\includegraphics[width=1\linewidth]{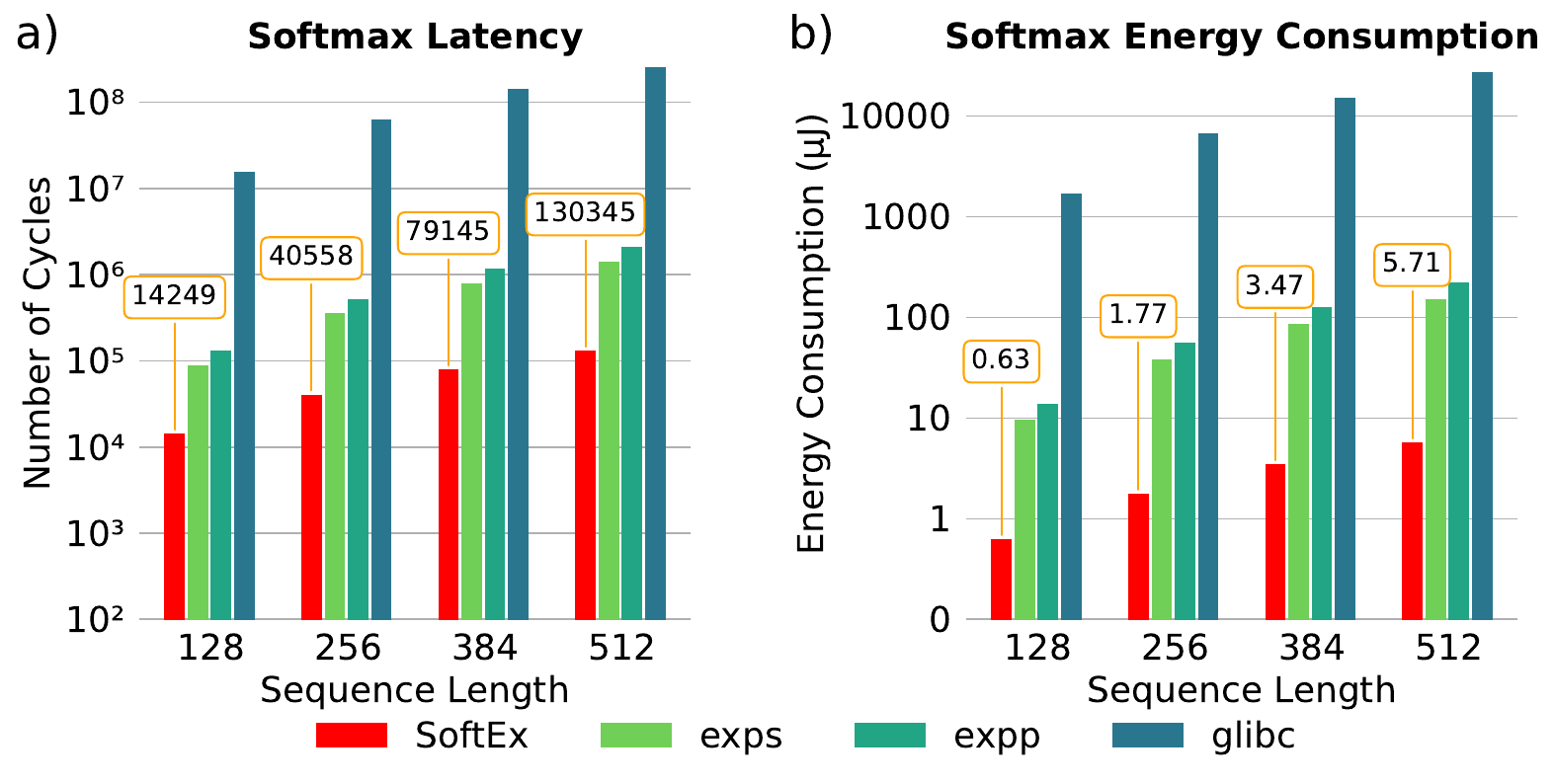}}
\caption{Comparison of SoftEx's latency (a) and energy consumption (b) at 0.80V with different software implementations.}
\label{softmax_perf}
\end{figure}

\begin{figure}[tbp]
\centerline{\includegraphics[width=1\linewidth]{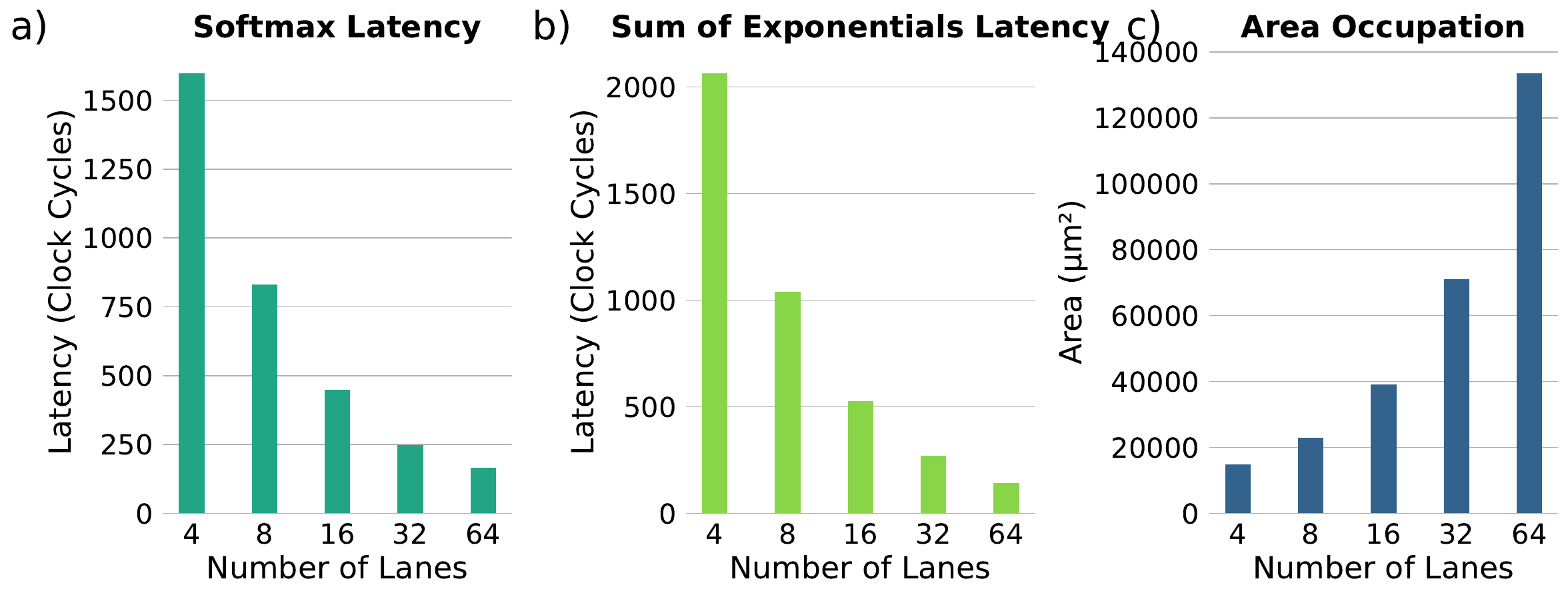}}
\caption{SoftEx's average latency on 2048-long vectors (a, b) and area (c) with different configurations.}
\label{softmax_sweep}
\end{figure}

\paragraph{Softmax performance analysis}
We benchmarked SoftEx softmax performance using activations from MobileBERT's attention layer with different sequence lengths and compared the results with three software implementations running on the RISC-V cores using different exponentiation algorithms: glibc's implementation, Schraudolph's method ($\mathrm{exps}$), and the algorithm described in Section~\ref{expu} ($\mathrm{expp}$).
All software implementations are parallelized to exploit the cluster's 8 RISC-V cores.
As shown in Fig.~\ref{softmax_perf}, even with a relatively short sequence length of~128, SoftEx outperforms the best-performing software approximation.
In this scenario, the various exponential implementations scale from contributing 15~Mcycles to the total softmax latency in the glibc case, down to 92.7~kcycles for $\mathrm{expp}$ and 51.2~kcycles for $\mathrm{exps}$, while for SoftEx they are included within the 14.2~kcycles of total latency.
Overall, the attention probabilities calculation using SoftEx is $6.2\times$ faster and requires $15.3\times$
%
%
%
less energy than the second best method ($\mathrm{exps}$).
If we increase the sequence length to 512, we see that the gap with software widens even further, with performance and energy gains over software of $10.8\times$ and $26.8\times$, respectively. 
We observe that on average $\mathrm{expp}$ results in a softmax only 31\% slower than $\mathrm{exps}$, highlighting the potential usefulness of our method even in a pure software implementation targeting high accuracy.

\begin{figure}[tbp]
\centerline{\includegraphics[width=0.85\linewidth]{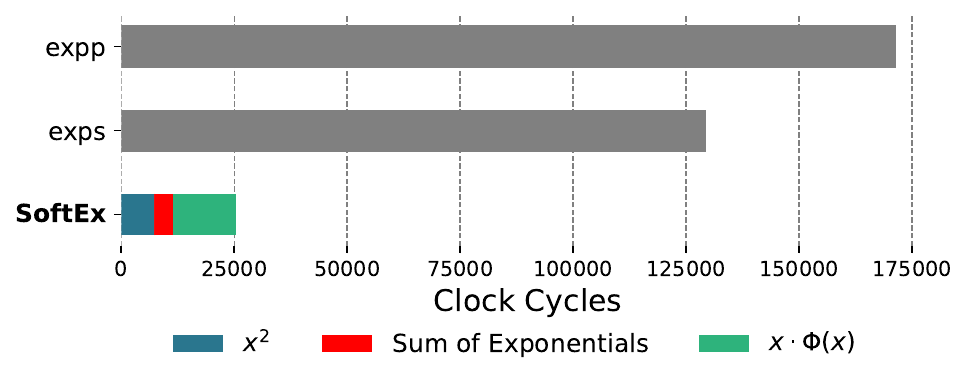}}
\caption{Runtime breakdown of GELU on $2^{14}$ elements calculated either using only the cores or assisted with SoftEx. When run entirely in software, GELU is approximated using the sigmoid function (Eq.~\ref{gelu_sigmoid}), when assisted with SoftEx, a four-term sum of exponentials is used.}
\label{gelu_perf}
\end{figure}

\paragraph{GELU performance analysis}
To verify the performance gains of using SoftEx in computing GELU, we benchmarked the system using activations from the first fully-connected network of a layer of ViT base.

As shown in Fig.~\ref{gelu_perf}, even if part of GELU is still performed in software, using SoftEx still results in a significant speedup over a full software implementation base on the sigmoid function. The SoftEx-assisted GELU achieves an overall speedup of $5.11\times$ and a $5.29\times$ energy reduction over the software-only implementation using $\mathrm{exps}$. If instead of Schraudolph's method we use the software implementation of the algorithm described in section \ref{expu} ($\mathrm{expp}$), thus achieving a comparable precision to that of the accelerator, the speedup and energy efficiency increase to $6.77\times$ and $7.02\times$ respectively.          

\begin{figure}[tbp]
\centerline{\includegraphics[width=0.9\linewidth]{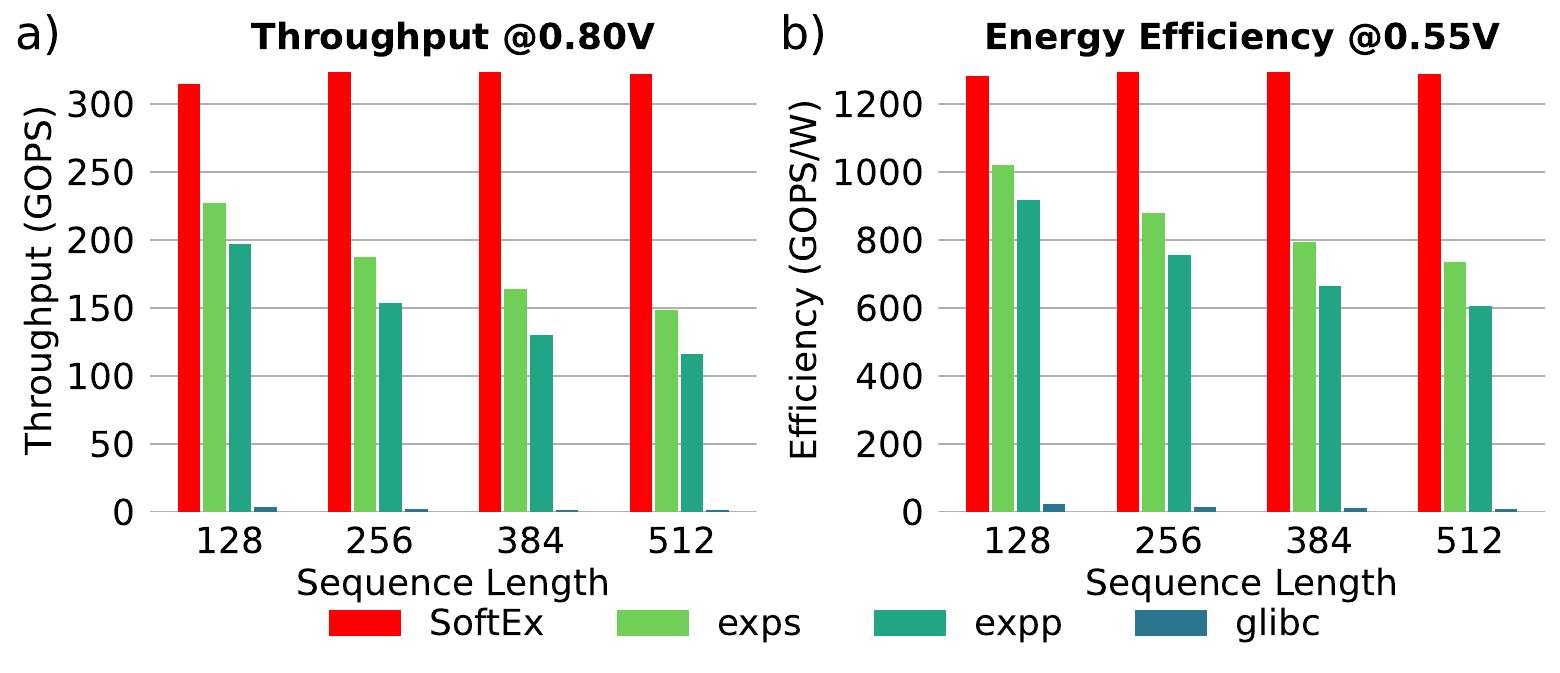}}
\caption{System throughput @0.80V (a) and energy efficiency @0.55V (b) on MobileBERT's attention layer using SoftEx and different software implementations.}
\label{system_perf}
\end{figure}

\begin{figure}[tbp]
\centerline{\includegraphics[width=1\linewidth]{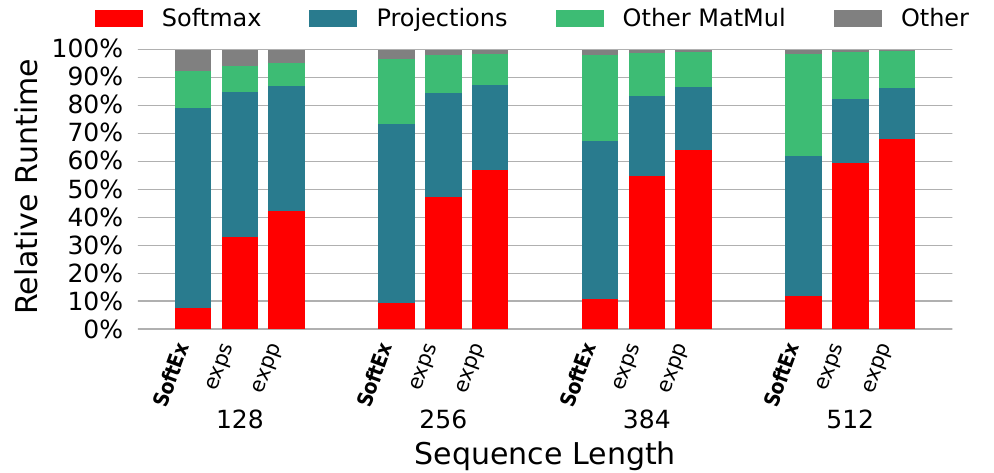}}
\caption{Runtime breakdown of the kernels inside MobileBERT's attention layer using SoftEx or the 8 cores (with $\mathrm{exps}$ or $\mathrm{expp}$) to calculate the softmax. In the glibc case (not shown), runtime is 99\% softmax-dominated.}
\label{softex_runtime}
\end{figure}

\paragraph{Performance and Area Scaling}
We instantiated SoftEx while sweeping the number of lanes and evaluated the impact on area and latency, benchmarked on vectors of 2048 elements. As shown in Fig.~\ref{softmax_sweep}, when the number of lanes is small, increasing them is extremely convenient. For example, changing the number of lanes from 4 to 8 almost doubles the performance in both softmax and the sum of exponentials, while increasing the accelerators's area by just 50\%. On the other hand, enlarging an already big accelerator has significantly lower returns, especially if the input vector is not much bigger than the available memory bandwidth, with the accelerator featuring 64 lanes being almost two times as large as the 32-lane version while being on average only 50\% faster in computing the softmax function. On the contrary, the sum of exponentials has a better scaling behavior: even when the bandwidth approaches the length of the input vector, its latency still scales linearly with the number of rows.
In conclusion, while the optimal lane configuration depends on the typical length of input vectors, a 16-lane design offers a good compromise, delivering significant performance improvements with a moderate increase in area, making it a balanced choice.

\begin{figure}[tbp]
\centerline{\includegraphics[width=0.85\linewidth]{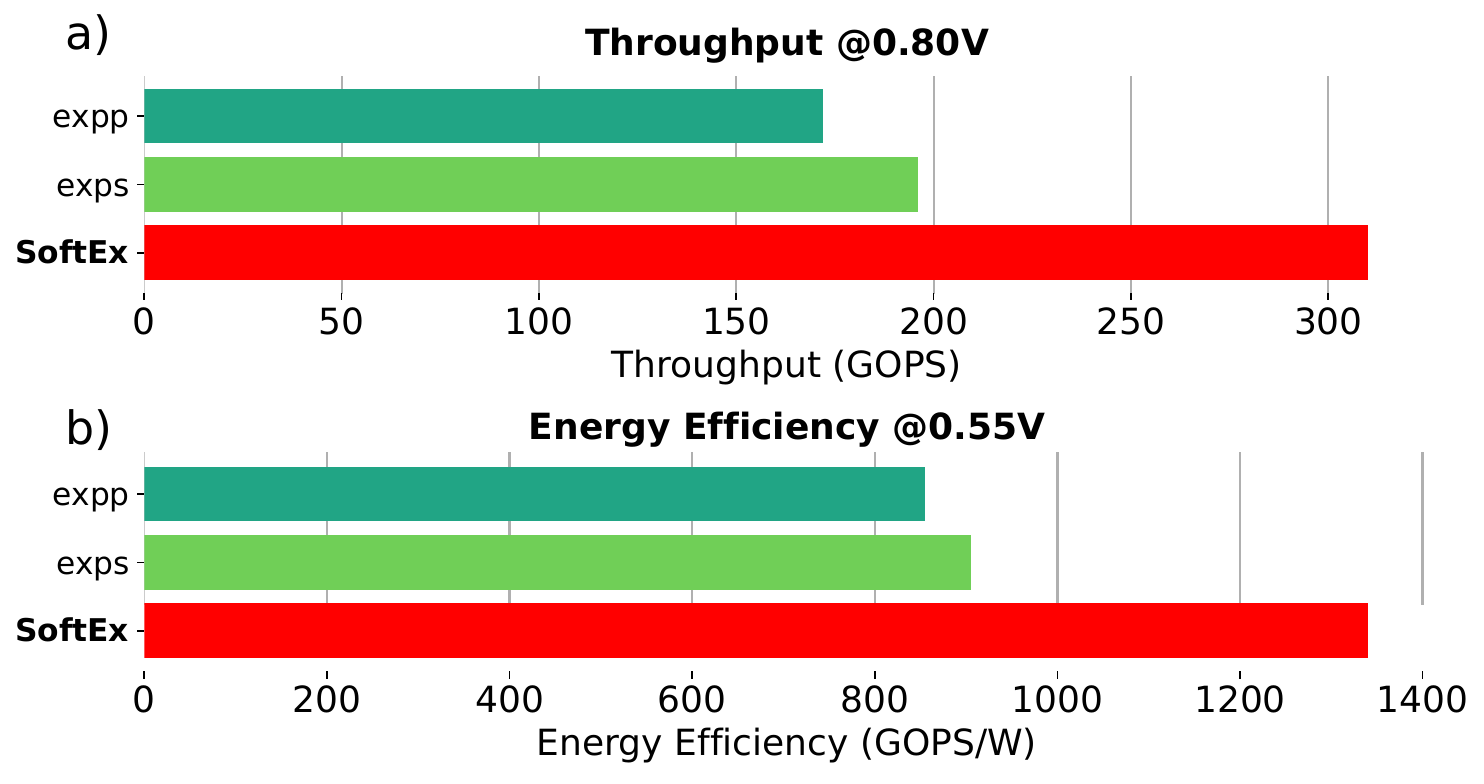}}
\caption{System throughput @0.80V (a) and energy efficiency @0.55V (b) on ViT using SoftEx and different software implementations.}
\label{vit_perf}
\end{figure}

\subsection{Cluster performance on MobileBERT}
Our proposed cluster employs a large tensor processing unit that achieves 430 GOPS at 0.8V and 1.72 TOPS/W at 0.55V, comparable with other similar units in the State-of-the-Art (see Section~\ref{sec:soa}) in the linear parts of the attention layer.
Here, we aim to assess \textit{1)} if the cluster is (as expected) affected by the softmax bottleneck; and \textit{2)} if the bottleneck is indeed alleviated by the proposed SoftEx accelerator.
To this end, we tested the performance of the system on MobileBERT's attention layer.
We show the results of this experiment in Figures \ref{system_perf} and \ref{softex_runtime}, which display the overall throughput/efficiency and the relative runtime distribution per kernel inside the attention subtile, respectively.

When performing the attention layer with SoftEx, at 0.8V the cluster achieves a compound throughput of up to 324 GOPS (75\% of the theoretical peak of purely linear operations).
As expected, all software implementations result in a substantial throughput loss ($>$2.17$\times$ slowdown for larger sequence sizes), even when targeting the fastest and least precise $\mathrm{exps}$. This is due to the large amount of time spent inside the softmax, as clearly visible in Fig.~\ref{softex_runtime}. In terms of energy efficiency, at 0.55V the system achieves up to 1.30 TOPS/W, improving this value by 20.5-75.4\% with respect to the eight RISC-V cores running Schraudolph $\mathrm{exps}$.
We conclude that in the tested configuration, SoftEx achieves its design goal of alleviating the softmax bottleneck and enables the cluster to achieve near-peak throughput and efficiency on a complete attention layer.

Moreover, we tested the proposed cluster on the 24 encoder layers of MobileBERT with a sequence length of 512, under the assumption of sufficient memory bandwidth on the external memory and using double buffering to hide the memory-related latencies. Including all the overheads introduced by other operators found in Transformers (e.g., the intermediate activations of the feed-forward neural networks), the cluster achieves an average throughput of up to 297 GOPS (69\% of the theoretical peak), with a total latency of 152 ms.

\begin{figure}[tbp]
\centerline{\includegraphics[width=0.85\linewidth]{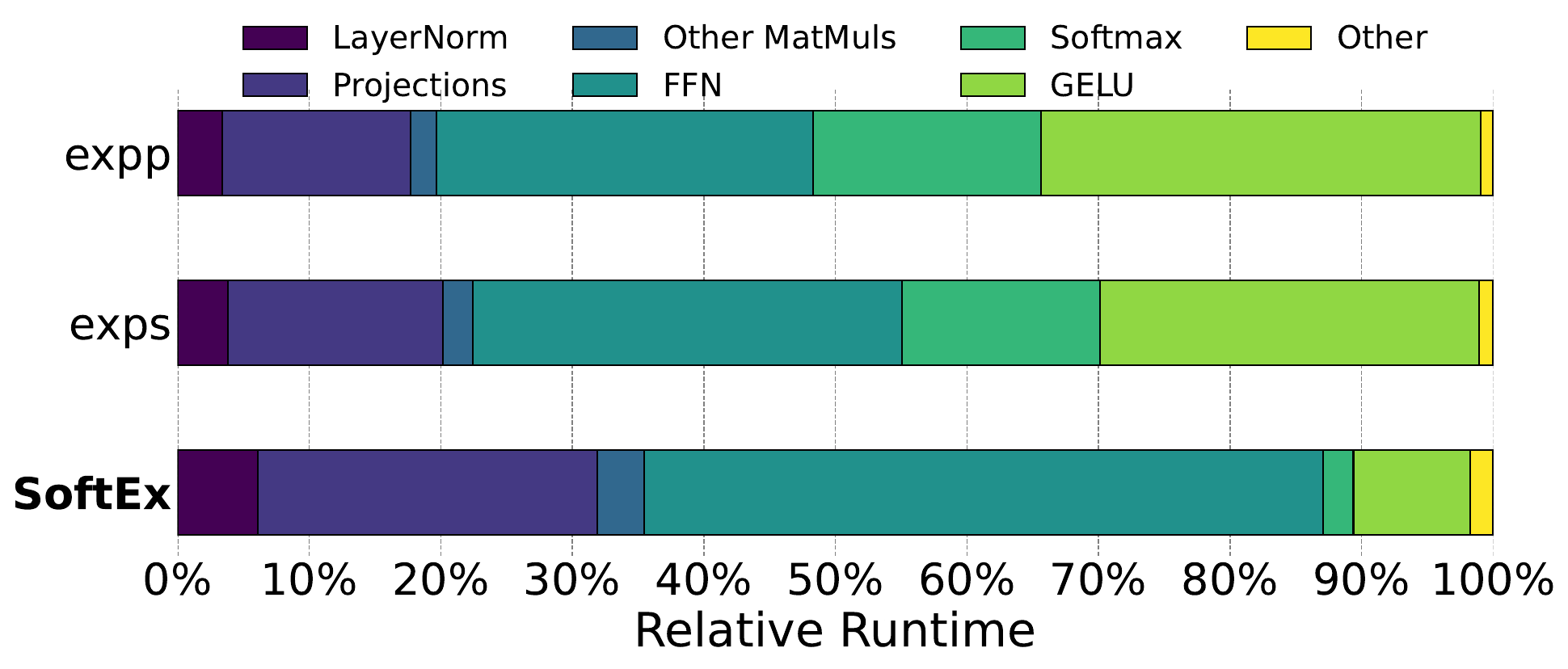}}
\caption{Runtime breakdown of the kernels inside ViT using SoftEx or the 8 cores (with $\mathrm{exps}$ or $\mathrm{expp}$) to calculate the softmax and GELU.}
\label{vit_runtime}
\end{figure}

\begin{figure}[tbp]
\centerline{\includegraphics[width=0.9\linewidth]{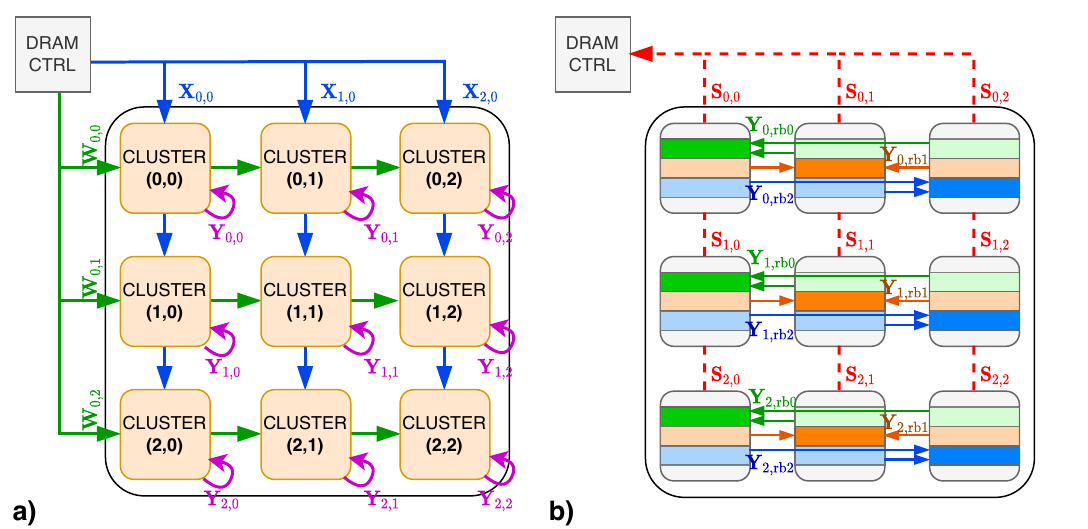}}
\caption{Dataflow modeled for scalability analysis. \textit{a)} $\mathbf{W}\cdot\mathbf{X}$ MatMul phase: the clusters share work in a systolic fashion with stationary output $\mathbf{Y}$; \textit{b)} softmax phase: each cluster collects a row-block (rb) from the $\mathbf{Y}$ of neighbour clusters in the horizontal direction to compute their part of the softmax $\mathbf{S}$.}
\label{fig:scaleup_mesh}
\end{figure}

\begin{figure*}[tbp]
\centerline{\includegraphics[width=0.9\linewidth]{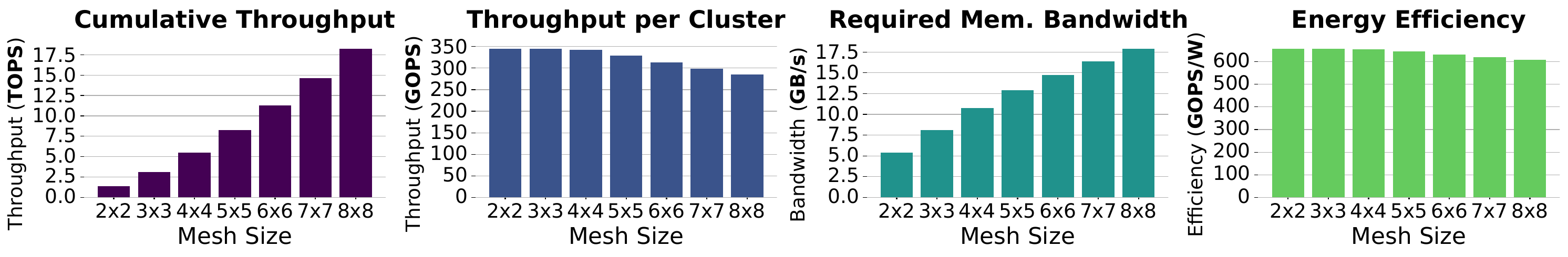}}
\caption{Performance figures of meshes of clusters of different sizes on GPT-2 XL. From left to right: The cumulative throughput achieved by the mesh of clusters; The average throughput of the single clusters; The bandwidth required by the system; The energy efficiency of the mesh in the nominal 0.8V operating point.}
\label{mesh_perf}
\end{figure*}

\subsection{Cluster performance on ViT}
To verify whether SoftEx also effectively mitigates the GELU bottleneck, we tested its performance on the Vision Transformer (ViT), which in the base configuration features a fixed sequence length of 197, an embedding size of 768, and a multi-headed attention mechanism with 12 heads.
While its main role is perceptive as part of vision pipelines, ViT is also a relevant part of many GenAI pipelines, e.g. BEiT~\cite{baoBEiTBERTPreTraining2022}.
Each layer’s feed-forward neural network consists of a linear transformation producing 3072 output dimensions, followed by a GELU nonlinearity, and another linear transformation reducing the dimensions back to 768.

As shown in Fig.~\ref{vit_runtime}, under this setup, when both Softmax and GELU are fully executed on the cores (with GELU approximated using the sigmoid approximation and Schraudolph's method), GELU becomes the primary bottleneck (28.8\% of the total), while Softmax contributes significantly less to the overall runtime (15.1\%).

At 0.8V enabling SoftEx boosts the cluster throughput to 310 GOPS (72\% of the theoretical peak), representing a 1.58$\times$ increase compared to software-only activations. The SoftEx-enhanced cluster achieves an end-to-end latency of 113 ms.
From an energy efficiency viewpoint, at 0.55V, the test system with SoftEx enabled achieves 1.34 TOPS/W, a $1.42\times$ improvement over the approximate software implementation.
In conclusion, even in models where the bottleneck is shared across multiple activations, SoftEx effectively achieves its purpose of alleviating the inefficiencies introduced by common nonlinearities, delivering substantial performance and energy efficiency gains at the cluster level.

\section{Scalability Analysis}
\label{sec:scalability}

To determine whether SoftEx is useful for accelerating large transformer models, we modeled the integration of various SoftEx-augmented clusters in a compute mesh based on FlooNoC~\cite{fischerFlooNoC645Gbps2024} and performed a scalability analysis on GPT-2 XL~\cite{radfordLanguageModelsAre2019}.
FlooNoC is a high-performance Network-on-Chip (NoC) architecture with computing elements arranged in a 2D-mesh. It features efficient (0.15pJ/B/hop) high-bandwidth AXI4 links, comprising two narrow 64-bit channels for latency-sensitive transactions and a wide 512-bit channel for high-bandwidth, latency-insensitive traffic. 

The clusters are organized in a $n \times n$ mesh configuration.
We split the workload with the strategy presented in Fig.~\ref{fig:scaleup_mesh}. 
During MatMuls, we split the matrices in square tiles and employ an output-stationary systolic dataflow, where each cluster retains outputs and propagates inputs to adjacent units (Fig.~\ref{fig:scaleup_mesh}a).
We keep data stationary to compute all pointwise non-linearities.
We use double buffering to hide the latency introduced by memory accesses and load data in chunks of 16K elements / 32KB.
During the computation of softmax, we marshal data so that each cluster can compute on one or more full rows: each cluster collects a row-block from contributions of adjacent clusters, with horizontal cluster-to-cluster communication (Fig.~\ref{fig:scaleup_mesh}b).
We assume all data to be ultimately accessed from/written back to an external DRAM.
To ease the modeling of the system, we made the following conservative assumptions: \textit{i)}  every conflict is independent; \textit{ii)} due to conflicts, each hop introduces an additional delay, which is given by the value of a random variable characterized by a uniform distribution in the [0,0.5] range (expressed in clock cycles per transaction); \textit{iii)} the overall additional delay on the execution of the network is given by the maximum total delay among all paths starting from the top-left tile and ending in the bottom-right tile.
We modeled the simplified graph in the NetworkX Python library and calculated the delays with a Monte Carlo procedure with $2^{16}$ trials each.

In this configuration, on GPT-2 XL in prompt mode, the tensor processing unit utilization is on average 80\%, translating to a maximum achievable performance per cluster of 345 GOPS. At full data path utilization, each router manages two inbound streams and two outbound streams. With the 512-bit wide AXI bus employed in the NoC, transferring four 32KB packets takes 2048 cycles, 16.9\% of the average time a cluster takes to process a chunk.

As seen in Fig.~\ref{mesh_perf}, the largest configuration ($8\times8$) achieves a throughput of 18.2 TOPS, $52.8\times$ higher than the performance of a single cluster. 
While the interconnect causes almost no overheads on meshes smaller than $4\times4$ tiles,  starting from the $5\times5$ configuration, the conflicts begin to have a significative impact on the overall performance, reaching a slowdown of up to 17.4\% on the largest configuration.
The energy efficiency follows a trend similar to the throughput, but with a less pronounced decline in larger configurations, with the $8\times8$ mesh being only 7.44\% less efficient than a single cluster.
Even in the $8\times8$ configuration, the contribution of the NoC on the overall power consumption is minimal, accounting for just 0.29\% of the total.
The aggregate bandwidth requirement scales sub-linearly, with the smallest and bigger configurations requiring to be fed with 5.42 GB/s and 17.9 GB/s of data respectively. 
In terms of pure bandwidth, even the largest mesh could be fed with a single 6400 MT/s LPDDR5 DRAM~\cite{LPDDR5PartDetail}.

\section{Comparison with the State of the Art} 
\label{sec:soa}

\subsection{Comparison to other softmax accelerators}

\begin{table*}[tb]
\caption{Comparison of the test system to different state-of-the-art transformer accelerators}
\begin{center}
\resizebox{0.95\textwidth}{!}{
\begin{tabular}{l c c c c c c c c c c c c}
     & & {Tambe et al. \cite{tambe22912nm181TFLOPs2023}}    & & ITA \cite{wieseAttentionbasedTinyMLHeterogeneous2024}       & & {Keller et al.\cite{keller17956TOPSDeep2022}} & & ViTA \cite{chenViTAHighlyEfficient2024} & & Dumoulin et al.\cite{dumoulinEnablingEfficientHardware2024} & & \textbf{This Work} \\
    \cline{3-3} \cline{5-5} \cline{7-7} \cline{9-9} \cline{11-11} \cline{13-13} \\
    \textbf{Data Format}                  & & FP8                   & & INT8                        & & INT8                        & & INT8                         & & INT8    & & BF16 \\
    \textbf{Technology} (nm)              & & 12                        & & 22                          & & 5                           & & 28                       & & 28  & & 12   \\
    \textbf{Area} (mm$^2$)                & & 4.60                  & & 0.991                       & & 0.153                       & & 2.00                        & & 1.48    & & 1.21 \\
    \textbf{Voltage} (V)                  & & 0.62-1.0              & & 0.65                         & & 0.46-1.05                   & & 1.05                    & & -   & & 0.55-0.8  \\
    \textbf{Power} (mW)                   & & 10-122                & & 132                         & & -                           & & 217                     & & 18.4    & & 110-581  \\
    \textbf{Frequency} (MHz)              & & 77-717                & & 425                         & & 152-1760                    & & 200                    & & 100  & & 460-1120 \\
    \textbf{MAC Units}                    & & 256                   & & 1024                        & & 512                         & & 512              & & 256   & & 192  \\
    \textbf{On-Chip SRAM} (KiB)           & & 647                   & & 128                        & & 141                         & & 48                & & 512   & & 256  \\
    \textbf{Supported Nonlinearities} && Softmax &&Softmax &&Softmax &&Softmax, GELU && Softmax && Softmax, GELU \\
    \textbf{Peak Throughput} (GOPS)            & & 367                 & & 870                        & & 1800                         & & 204                  & & 51.2             & & 430 \\
    \textbf{Peak Energy Efficiency} (TOPS/W) & & 3.0                 & & 5.49                        & & 39.1$^{*}$                        & & 0.943            & & 2.78    & & 1.61 \\
    \hline
\end{tabular}
}
\label{tab:SOA}
\end{center}
$^{*}$ Assuming 50\% of the inputs is zero
\end{table*}

With a 10 16-bit fixed-point inputs configuration, Zhang~et~al.~\cite{zhangBase2SoftmaxFunction2022} report an operating frequency of 3.0 GHz with a power consumption of 13.12 mW for their accelerator, implemented in 28 nm technology and configured with 10 16-bit fixed-point inputs. On the other hand, their design assumes the entire input vector fits within the available bandwidth, making it unsuitable when the maximum input length is unknown a priori and requiring an exceptionally high bandwidth for larger inputs.

Zhu~et~al.~\cite{zhuEfficientPrecisionAdjustableArchitecture2020} also implemented in 28 nm technology, achieve maximum frequencies of 2.78 GHz and 1.64 GHz for their less precise and more accurate softmax implementations, respectively. However, they do not provide information on the operating conditions nor the average power consumption. Their approach also relies on a naive softmax implementation requiring three passes over the input vector, significantly limiting its efficiency.

Kim~et~al.~\cite{kimHardwareEfficientSoftMaxArchitecture2024} proposed a fixed-point softmax accelerator that requires a single pass on the inputs. Implemented in 28 nm technology and operating at 1.0V, their design achieves a maximum frequency of 2.5 GHz with an average power consumption of 52.46 mW for an instance processing 8 16-bit inputs per cycle. However, the design has two critical limitations: it requires the maximum input value to be provided externally and necessitates internal buffering of all inputs, restricting its applicability to scenarios where the input length aligns with the buffer capacity.

The above-mentioned accelerators report remarkable operating frequencies and performance. On the other hand, these results are extracted from a post-synthesis simulation of the accelerators only, not accounting for the additional overheads introduced by the integration in a system, making a meaningful comparison impractical since in real systems the critical path is often through the memories or other compute components of the system (e.g. processor pipeline, interconnects, etc...).


\subsection{Comparison to other Transformer accelerators}
We compare our design with State-of-the-Art Transformer accelerators in Table~\ref{tab:SOA}.
Compared to the accelerator proposed by Tambe~et~al.\cite{tambe22912nm181TFLOPs2023}, our cluster achieves a 17\% higher peak theoretical throughput (430 GOPS vs. 367 GOPS) while attaining 54\% of their peak energy efficiency (1.61 TOPS/W vs. 3.0 TOPS/W). Notably, these results are achieved despite Tambe~et~al.'s design utilizing the less precise FP8 format and incorporating 256 MAC units, 33\% more than our tensor processing unit.

In its smaller and more efficient configuration, which features an area comparable to the proposed cluster, at 200 MHz ViTA~\cite{chenViTAHighlyEfficient2024} achieves a maximum theoretical throughput of 204 GOPS, 52.6\% lower than that of our work at 0.8V, and an efficiency of 0.943 TOPS/W , 42\% lower than out cluster at 0.55V.
While ViTA supports a broader range of non-linear functions thanks to the integration of the algorithm proposed in \cite{yuNNLUTNeuralApproximation2022}, this comes at the cost of an area which is $1.29\times$ the one of the Vector-Matrix units. In contrast, while we support less functions, SoftEx's area is just 16.5\% of that of the tensor processing unit.
At 300MHz, ViTA is reported to achieve an end-to-end latency of 91 ms on the base Vision Transformer, just 19\% faster than that of the proposed cluster and achieved in the narrower INT8 format. Moreover, using integer formats requires the model to be quantized and fine-tuned for the less precise numerical format format before being run on the accelerator.
The ITA-enhanced cluster presented in \cite{wieseAttentionbasedTinyMLHeterogeneous2024} has a peak theoretical throughput that is more than twice ours, albeit achieved using the INT8 format and employing 5.33$\times$ more MAC units than our design. For the attention layer specifically, ITA attains a peak throughput of 620 GOPS, which is 1.91$\times$ higher than our performance on MobileBERT's attention layer. However, our cluster achieves a tensor processing unit utilization of 75\%, 5.5\% higher than that of ITA’s system.
%
%
Despite the less advanced technology node, at a power consumption of just 18.4 mW Dumoulin~et~al.~\cite{dumoulinEnablingEfficientHardware2024} achieve an energy efficiency of 2.78 TOPS/W, 73\% better than ours. This comes at the cost of performance, as their throughput is only 51.2 GOPS, 8.40$\times$ and 3.45$\times$ lower than our peak throughput at 0.8V and 0.55V respectively. 

Keller et al.~\cite{keller17956TOPSDeep2022} report a $29\times$ higher efficiency compared to our cluster;  however this is achieved using a more advanced technology node, relying on an 8-bit integers and under the very strong assumption of 50\% sparsity. In contrast, our design makes no assumption on the model that is being executed and requires no intervention on the networks.

\subsection{Comparison to large SoCs}
\begin{table}[tb]
\begin{center}
\resizebox{0.9\columnwidth}{!}{
\begin{threeparttable}
\caption{Comparison with academic and commercial SoCs}
\label{tab:soa_socs}
\begin{tabular}{@{}lll@{}}
\toprule
\multirow{2}{*}{\textbf{Architecture}}                         & \textbf{Performance (BF16)} & \textbf{Efficiency (BF16)} \\
                         & {[}TOPS{]}           & {[}TOPS/W{]}        \\ \midrule
\textbf{Our work}, 8$\times$8 mesh (12nm) & \textbf{18.20}       & \textbf{0.60}       \\
Occamy (12nm)            & 0.72                 & 0.15                \\ \midrule
\textbf{Our work}, 8$\times$8 mesh (7nm)$^{*}$ & \textbf{18.20}       & \textbf{1.56}       \\
Occamy (7nm)$^{*}$            & 0.72                 & 0.39                \\
Nvidia A100 (7nm)        & 312.00               & 1.04                \\
\bottomrule
\end{tabular}
\begin{tablenotes}
\item$^{*}$Scaled with $P_{7nm} = P_{12nm} \cdot (7/12) \cdot (V_{7nm}/V_{12nm})^2$
\end{tablenotes}
\end{threeparttable}
}
\end{center}
\end{table}
To position our work in a larger context, in Table~\ref{tab:soa_socs} we compare the up-scaled mesh of clusters modeled in Section~\ref{sec:scalability} with two state-of-the-art SoCs that can be used for Transformer acceleration: an academic one, Occamy~\cite{paulinOccamy432Core2812024,potocnikOptimizingFoundationModel2024}, and NVIDIA's A100\footnote{\url{https://www.nvidia.com/en-us/data-center/a100/}}.
Occamy and the 8$\times$8 mesh occupy similar amounts of silicon area: 51.5mm$^\text{2}$ for Occamy's compute regions, and $\sim$80mm$^\text{2}$ for our mesh ($\sim$1.6$\times$ more).
Despite this, our mesh would provide $\sim$25$\times$ the performance at 4$\times$ better BF16 efficiency, at the cost of flexibility.
Compared to NVIDIA A100, despite the different targets (edge AI vs HPC), our mesh scaled to the same technology node would be competitive in BF16 efficiency to TensorCores, providing a $\sim$50\% boost compared to that reported by NVIDIA.

\section{Conclusion}
\label{sec:conclusion}
We presented a flexible acceleration template for GenAI Transformers at the edge, based on a 8-core RISC-V cluster augmented with a $24\times8$-PEs tensor processing unit and with SoftEx, a novel accelerator for BFloat16 softmax and GELU non-linearities. 
Thanks to SoftEx, which relieves the bottleneck caused by non-linearities when MatMul is accelerated, our template achieves up to 310 GOPS (72\% of the theoretical peak) at 0.80V or an energy efficiency of up to 1.34 TOPS/W at 0.55V on ViT. 


\bibliographystyle{IEEEtran}
\bibliography{ieee,main_without_url,main_withurl}

\newpage

\section*{Appendix: GELU approximation}

In this work, we calculate the $\Phi$ function using an approximation of the closely related Q function.
The Q function is defined as the complementary cumulative distribution function of a Gaussian random variable:
\begin{equation}\label{q_func}
    \begin{aligned}
    Q(x) &\doteq 1 - \Phi(x) = \mathrm{Prob}(X \geq x) \\
         &= \frac{1}{\sqrt{2\pi}} \int_x^\infty \exp\big(-\frac{1}{2}t^2\big)dt
    \end{aligned}
\end{equation}
For positive arguments, it can be shown that this function can be expressed as:
\begin{equation}\label{q_craig}
    Q(x) = \frac{1}{\pi} \int_0^{\frac{\pi}{2}}\exp\left(-\frac{x^2}{2\sin^2\theta}\right)d\theta, x\geq0
\end{equation}

Starting from this definition, Chiani et al. \cite{chianiNewExponentialBounds2003} note that, as the integrand is monotonically increasing, the rectangular integration rule can be applied to obtain a tight upper bound of the Q function:
\begin{equation}\label{q_chiani_appendix}
    \begin{aligned}
        Q(x) &\leq \frac{1}{\pi} \sum_{i=1}^N\int_{\theta_{i-1}}^{\theta_i}\exp\left(-\frac{x^2}{2\sin^2\theta_i}\right)d\theta \\
             &=    \sum_{i=1}^N a_i e^{-b_i x^2}, x\geq0
    \end{aligned}
\end{equation}
Building on this result, Tanash and Riihonen \cite{tanashGlobalMinimaxApproximations2020} propose a method to calculate the $a$ and $b$ parameters such that the resulting approximation is optimal in a minmax sense. 
The best minmax approximation occurs when the error function oscillates uniformly between equal-magnitude extrema, so the optimal $a$,$b$ parameters, alongside the $2N$ maximum error points $x_k$ and the maximum absolute error $d_\mathrm{max}$, are determined by solving:
\begin{equation}\label{q_final}
    \begin{aligned}
        \begin{cases}
            d'(x_k) = 0, & \quad \mathrm{for} \; k = 1,2,...,2N, \\
            d(x_k) = (-1)^{k+1}d_\mathrm{max}, & \quad \mathrm{for} \; k = 1,2,...,2N, \\
                \sum_{n=1}^{N}a_n = \frac{1}{2}, & \quad \mathrm{when} \; d(0) = 0, \\
                \sum_{n=1}^{N}a_n = \frac{1}{2} - d_{\mathrm{max}}, & \quad \mathrm{when} \; d(0) = -d_{\mathrm{max}}
        \end{cases}
    \end{aligned}
\end{equation}

To set up the problem in terms of relative error, Tanash and Riihonen first note that the relative error $r(x)$ does not converge to zero as $x\to\infty$. Therefore,  the function has to be minimized over a finite range $[0, x_{2N+1}]$, where $x_{2N+1}$ is an arbitrarily chosen point at which $r(x_{2N+1})=-r_{\mathrm{max}}$ before the relative error asymptotically approaches -1.
Therefore, the optimal values of $a$ and $b$, the $2N$ maximum error points $x_k$, and the maximum relative error $r_{\mathrm{max}}$ are found by solving:
\begin{equation}\label{q_final_relative}
    \begin{aligned}
        \begin{cases}
            r'(x_k) = 0, & \mathrm{for} \; k = 1,2,...,2N, \\
            r(x_k) = (-1)^{k+1}r_\mathrm{max}, & \mathrm{for} \; k = 1,2,...,2N, \\
                \sum_{n=1}^{N}a_n = \frac{1}{2}, & \mathrm{when} \; r(0) = 0, \\
                \sum_{n=1}^{N}a_n = \frac{1}{2} - \frac{r_{\mathrm{max}}}{2}, & \mathrm{when} \; r(0) = -r_{\mathrm{max}} \\
            r(x_{2N+1}) = -r_{\mathrm{max}}.
        \end{cases}
    \end{aligned}
\end{equation}

Using this sum-of-exponentials formulation, we approximate the Gaussian CDF with an adjustable precision and avoid expensive dividers. 
Moreover, as Craig's formulation is symmetric, it can be shown that, for $x < 0$, Eq. \ref{q_craig} and, consequently, the result of Eq. \ref{q_final},\ref{q_final_relative}, compute $\Phi(x)$ rather than $Q(x)$.
The GELU function can thus be implemented as a four-step procedure: \textit{1)} square the input $x$; \textit{2)} use the result of \textit{1)} to calculate $\sum_{i=1}^N a_i e^{-b_i x^2}$; \textit{3)} if $x>0$, complement the result of \textit{2)}; \textit{4)} multiply $x$ by the result of \textit{3)}.

\end{document}